%% file: main.tex
\documentstyle[11pt,epsf]{article}
\include{style}
\include{symb}

\title{Semiclassical trace formulas in terms of phase space path
integrals}
\author{Ayumu Sugita\\
Department of Physics,Graduate School of Science,Kyoto University,\\
Kyoto 606-01,Japan}
\date{}

\begin{document}
\maketitle
\include{abs}

\newpage
\tableofcontents
\newpage
\include{intro}

\include{SPA}

\include{quad}

\include{g}

\include{concl}

\include{bib}
\appendix
\include{diag}
\include{direct}

\include{formula}

\end{document}

%% file: symb.tex
\newcommand{\tr} {\mbox{${\rm Tr}$}}
\newcommand{\bp} {\mbox{\boldmath$p$}}
\newcommand{\bq} {\mbox{\boldmath$q$}}
\newcommand{\bv} {\mbox{\boldmath$v$}}
\newcommand{\bx} {\mbox{\boldmath$x$}}
\newcommand{\by} {\mbox{\boldmath$y$}}
\newcommand{\bX} {\mbox{\boldmath$X$}}
\newcommand{\bxi}{\mbox{\boldmath$\xi$}}

\newcommand{\bm} {\mbox{\boldmath$m$}}

\newcommand{\Det}    {\mbox{${\rm Det}$}}
\newcommand{\Dslash} {\mbox{$D\hskip -0.65em /\;$}}
\newcommand{\Gsim}      {\mbox{$\;\stackrel{G}{\sim}\;$}}
\newcommand{\Lsim}      {\mbox{$\;\stackrel{L}{\sim}\;$}}
\newcommand{\Gzerosim}  {\mbox{$\;\stackrel{G_{0}}{\sim}\;$}}
\newcommand{\Lzerosim}  {\mbox{$\;\stackrel{L_{0}}{\sim}\;$}}

\newcommand{\eminus}    {\mbox{$e^{-i\frac{2\pi r}{N}}$}}

%% file: abs.tex
\abstract
Semiclassical trace formulas are examined using phase
space path integrals. Our main concern in this paper
is the Maslov index of the periodic orbit, which seems
not fully understood in previous works.
We show that the calculation of the Maslov index
is reduced to a classification of connections on a 
vector bundle over $S^{1}$ with structure group $Sp(2n,R)$. 
We derive a formula for the index of the n-repetition,
and show that a Bohr-Sommerfeld type quantization
condition including quadratic fluctuation around the
orbit is derived using this formula.

%% file: intro.tex
\section{Introduction}
The semiclassical trace formula developed 
by Gutzwiller \cite{gutzwiller} is one of the most important 
tools to study the spectrum of non-integrable systems.
However, the original derivation of this formula is 
very complicated, and the canonical invariance of 
the formula (especially the Maslov index) is unclear.

In this paper, we examine the semiclassical trace formula
using phase space path integral to clarify the canonical 
structure of this formula.
Our main concern here is the Maslov index, which is an
additional phase factor appearing in the trace formula.

Although many studies have been made on the geometrical 
properties of this index, it doesn't seem to be 
thoroughly understood. For hyperbolic orbits, Robbins 
\cite{robbins} showed that this index is equal to twice
the winding number which is defined by the invariant 
manifolds around the orbit. In this case, the Maslov 
index of the n-repetition of the orbit $\mu_{n}$ is 
equal to $n\mu_{1}$. However, Brack and Jain \cite{brack}
investigated periodic orbits in anisotropic harmonic 
oscillator (these orbits are elliptic), and showed 
that $\mu_{n}$ is not equal to $n\mu_{1}$.
This result suggests that we need new geometrical picture
to understand this index.

We start with the phase space path integral of the partition
function:
\begin{equation}
Z(T) = \int {\cal D}\bp{\cal D}\bq \exp 
\left[\frac{i}{\hbar}\oint (\bp d\bq - Hdt)\right].
\label{Z}
\end{equation}
The density of states is obtained as the Fourier-Laplace
transformation of the partition function:
\begin{equation}
\rho (E) = - \frac{1}{\pi}{\rm Im} g(E+ i\epsilon),
\end{equation}
\begin{eqnarray}
g(E) & = & {\rm Tr}\frac{1}{E-\hat{H}}, \\
& = & \frac{1}{i\hbar}\int_{0}^{\infty}dT e^{iET/\hbar}Z(T).
\end{eqnarray}
We obtain the semiclassical trace formula by applying the
stationary phase approximation to the path integral.

The stationary condition reads
\begin{equation}
\delta\oint (\bp d\bq - Hdt) = 0,
\label{stationary}
\end{equation}
which leads to the Hamiltonian equation of motion.
Since the periodic boundary condition of the path integral 
(\ref{Z}) has the effect of closing the paths, the solutions
of (\ref{stationary}) are periodic orbits.

Thus the partition function is approximated by a sum over 
the periodic orbits:
\begin{equation}
Z(T) = \sum_{p.o} K \exp \left[\frac{i}{\hbar}R\right].
\end{equation}
Here, $R=\oint \bp d\bq - Hdt$ is the classical action of 
the periodic orbit, and $K$ is the contribution of the
quadratic fluctuation around the orbit:
\begin{equation}
K = \int {\cal D}\bx \exp [i\delta^{2}R[\bx (t)]],
\end{equation}
\begin{equation}
\bx (t) = \frac{1}{\sqrt{\hbar}} (\delta\bq, \delta\bq).
\end{equation}
The Maslov index appears as the phase factor of this 
quadratic path integral. Since this is a kind of 
Fresnel integral, the phase is determined by the 
signs of diagonal elements of the quadratic form
$\delta^{2}R$.

Strictly speaking, we should write (\ref{Z}) in a discrete
form to obtain well-defined continuum limit. 
Since the way of the discretization reflects the operator ordering
of the Hamiltonian, we should treat this problem carefully 
\cite{kashiwa}. 
In this paper, we always use the mid-point prescription, which
corresponds to the Weyl-ordering of the operators.
We can formulate the canonical transformations of the path integral
most clearly in this prescription, as we will see in \S \ref{quadra}.

\begin{figure}
\epsfxsize = 0.6\textwidth
\centerline{\epsffile{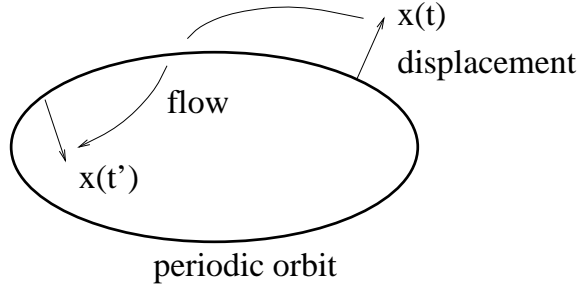}}
\caption{The set of displacement vectors is regarded as 
a fibre bundle, and Hamiltonian symplectic flow defines the
connection. The structure group of this space is $Sp(2n,R)$}.
\label{bundle}
\end{figure}

The central idea of this paper is that the Maslov index of the
periodic orbit is determined by the linearized symplectic flow 
around the orbit. The set of displacement vectors $\{\bx(t)\}$
is considered to be a vector bundle over $S^{1}$, and the flow 
around the orbit define a connection (Fig. \ref{bundle}).
Mathematically speaking, our problem is the classification of 
connections on the fibre bundle. However, the quadratic path integral around
the periodic orbit is not invariant to all canonical 
transformations. If the canonical transformation is topologically
non-trivial, the path integral may change the sign. This is 
essentially the same as
the global anomalies of the gauge field theories 
\cite{witten,elitzure}.
Therefore we regard two connections as equivalent if they
are connected by a topologically trivial canonical transformation,
and classify the connections by this equivalence relation. 

This paper is organized as follows.
In \S \ref{path}, we review the derivation of the phase
space path integral of the partition function and the 
stationary phase approximation. Note that the discussion 
in this section is not restricted to the time-independent
Hamiltonians.
\S \ref{quadra} is the main
part of this paper, and we treat the quadratic path integrals
with periodic boundary conditions generally. We calculate the
path integrals by reducing them to simple normal forms using
canonical (gauge) transformations.
We also derive a formula for the Maslov index of the n-repetition
of the orbit. In \S \ref{g}, we discuss the trace of the resolvent
$g(E)$ for time-independent systems. If the Hamiltonian of the system
is time-independent, the periodic orbit in it 
have at least one zero-mode. Therefore we first discuss the integration
with respect to zero-modes. Then we execute the Fourier
transformation of the partition function and derive the semiclassical
approximation to $g(E)$. We also show that the sum over repetitions
of a orbit leads to a Bohr-Sommerfeld type quantization condition.
Similar attempts have been made by many authors, but our result is the
first one which includes the effect of the Maslov index correctly.

%% file: SPA.tex
\section{Path integral and stationary phase approximation}\label{path}

\subsection{Path integral representation of the partition function}
In this subsection, we derive phase space path integral of the partition
function
\begin{equation}
Z(T) = \tr \hat{U}(T).
\end{equation}
Here,
$\hat{U}(T)$ is the time evolution operator 
\begin{equation}
\hat{U}(T) = {\rm P}\exp \left[\frac{i}{\hbar}
\int_{0}^{T}dt \hat{H}(t)\right],
\end{equation}
and $\hat{H}(t)$ is the Hamiltonian of the system. ${\rm P}$ denotes
the path ordered (or time ordered) product.

$\hat{U}(T)$ can be decomposed into a product of infinitesimal time
evolution operators:
\begin{eqnarray}
\hat{U}(T) & = & \prod_{j=1}^{N}
\left\{1-\frac{i}{\hbar}\Delta t\hat{H}(t_{j})\right\} + O(1/N),
\end{eqnarray}  
where $\Delta t$ denotes $T/N$,and $t_{j}$ denotes $j\Delta t$. 
Therefore $Z(T)$ can be written as
\begin{eqnarray}
Z(T) & = & \int \! d\bq \,\langle \bq | \hat{U}(T)| \bq\rangle, \\
& = & \lim_{N\rightarrow\infty}
\int \! \prod_{j=1}^{N}d\bq_{j} \langle \bq_{j}|
1-\frac{i}{\hbar}\Delta t\hat{H}(t_{j}) |\bq_{j-1}\rangle.
\end{eqnarray}
Here, the variables $\{q_{j}\}$ are cyclic and $q_{i}$ is considered to
be the same as $q_{j}$ if $i\equiv j ({\rm mod} N)$. This notation is
used throughout this paper.

The Feynmann kernel for the infinitesimal time interval
can be written as
\begin{eqnarray}
\langle \bq_{j} |1-\frac{i}{\hbar}\Delta t\hat{H}(t_{j}) |\bq_{j-1}\rangle
&=& \int \frac{d\bp_{j}}{(2\pi\hbar)^{n}} 
    \left\{1 - \frac{i}{\hbar}\Delta t 
    H_{W}\left(\bp_{j},\frac{\bq_{j}+\bq_{j-1}}{2},t_{j}\right)\right\}
    e^{i\bp_{j}(\bq_{j}-\bq_{j-1})/\hbar}, \\
&\simeq& \int \frac{d\bp_{j}}{(2\pi\hbar)^{n}} \exp\frac{i}{\hbar}
    \left[\bp_{j}(\bq_{j}-\bq_{j-1}) 
     - H_{W}\left(\bp_{j},\frac{\bq_{j}+\bq_{j-1}}{2},t_{j}
\right)\Delta t\right],  
\end{eqnarray}
where $n$ is the number of degrees of freedom, and 
$H_{W}$ is Weyl transform of $\hat{H}$:
\begin{equation}
H_{W}(\bp,\bq) = 
\int \! d\bv e^{i\bp\bv/\hbar}
\langle \bq-\frac{\bv}{2}|\hat{H}|\bq+\frac{\bv}{2}\rangle.
\end{equation}
Therefore we obtain the phase space path integral representation
of the partition function:
\begin{equation}
Z(T) = \lim_{N \rightarrow \infty} 
\prod_{j=1}^{N} \int \frac{d\bp_{j} d\bq_{j}}{(2\pi\hbar)^{n}}
\exp \left[\frac{i}{\hbar}\sum_{j=1}^{N}\left\{
\bp_{j}(\bq_{j}-\bq_{j-1}) - \Delta t H(\bp_{j},\overline{\bq}_{j},t_{j})
\right\}\right].
\label{pathint}
\end{equation}
Here, we denote $H_{W}$ as simply $H$, and 
$\overline{\bq}_{j}$ denotes $(\bq_{j}+\bq_{j-1})/2$
(mid-point prescription).

In the continuum limit, we denote this path integral as
\begin{equation}
Z(T) = 
\int {\cal D}\bq{\cal D}\bp 
\exp\left[\frac{i}{\hbar}\oint (\bp d\bq - Hdt)\right].
\label{pathintc}
\end{equation}
Note that the c-number Hamiltonian is defined by the Weyl
transform of the q-number Hamiltonian in this paper.
This c-number Hamiltonian can also be obtained by arranging 
the q-number Hamiltonian into Weyl ordering, and substituting
c-number variables for q-number variables \cite{kashiwa}. 

If the Hamiltonian is given as
\begin{equation}
\hat{H} = \frac{\hat{\bp}^{2}}{2m} + V(\hat{\bq},t),
\end{equation}
integration with respect to $\bp$ in (\ref{pathint}) can be
done easily, and we obtain the path integral representation 
of $Z$ in coordinate space:
\begin{equation}
Z(T) = \lim_{N\rightarrow\infty}\prod_{j=1}^{N}\int 
\left(\frac{m}{2\pi i\hbar\Delta t}\right)^{n/2}d\bq_{j}
\exp \left[\frac{i}{\hbar}\sum_{j=1}^{N}
\left\{\frac{m}{2}
\left(\frac{\bq_{j}-\bq_{j-1}}{\Delta t}\right)^{2} - 
V(\overline{\bq}_{j},t_{j})
\right\}\Delta t\right].
\label{lag1}
\end{equation} 
In the continuum limit, this formula can be denoted as
\begin{equation}
Z(T) = \int {\cal D}\bq \exp
\left[\frac{i}{\hbar}\oint Ldt\right],
\label{lag2}
\end{equation}
where $L$ is the Lagrangian of the system.

\subsection{Stationary phase approximation}
Let us evaluate the path integral (\ref{pathint}) by
the stationary phase approximation. We define a discretized action
$R_{N}$ as
\begin{equation}
R_{N}(\bp_{1},\bp_{2},..,\bp_{N},\bq_{1},..,\bq_{N}) = 
\sum_{j=1}^{N}\left\{
\bp_{j}(\bq_{j}-\bq_{j-1}) - \Delta t H(\bp_{j},\overline{\bq}_{j},t_{j})
\right\}.
\end{equation}
Then the stationary phase condition reads
\begin{eqnarray}
0 = \frac{\partial R_{N}}{\partial \bp_{j}}
& = &
\bq_{j} - \bq_{j-1} - \Delta t \frac{\partial H}{\partial \bp}
(\bp_{j},\overline{\bq}_{j},t_{j}),\\
0 = \frac{\partial R_{N}}{\partial \bq_{j}}
& = &
\bp_{j} - \bp_{j+1} - \frac{\Delta t}{2}
\left\{\frac{\partial H}{\partial\bq}
(\bp_{j+1},\overline{\bq}_{j+1},t_{j+1}) +
\frac{\partial H}{\partial \bq}(\bp_{j},\overline{\bq}_{j},t_{j})\right\},
\end{eqnarray}
which leads to the Hamiltonian equation of motion in the limit
$N\rightarrow \infty$:
\begin{eqnarray}
\dot{\bq} & = & \frac{\partial H}{\partial \bp}, \\
\dot{\bp} & = & - \frac{\partial H}{\partial \bq}.
\end{eqnarray}
Since variables in $R_{N}$ are cyclic, the solutions of the stationary
phase condition are classical periodic orbits.

The second derivatives of $R_{N}$ are
\begin{eqnarray}
\frac{\partial^{2}R_{N}}{\partial \bp_{i} \partial \bp_{j}}
& = &
- \Delta t(H_{pp})_{i,j},\\ 
\frac{\partial^{2}R_{N}}{\partial \bq_{i} \partial \bq_{j}}
& = &
- \Delta t (H_{qq})_{i,j},\\
\frac{\partial^{2}R_{N}}{\partial \bp_{i} \partial \bq_{j}}
& = &
\Delta t\{(\Delta)_{i,j} - (H_{pq})_{i,j}\}.
\end{eqnarray}
Here, $H_{pp},H_{pq},H_{qq}$ and $\Delta$ are $Nn\times Nn$ matrices,
which are defined as
\begin{eqnarray}
(H_{pp})_{i,j} 
& = & \frac{\partial^{2}H}{\partial \bp^{2}} 
(\bp_{j},\overline{\bq}_{j},t_{j})\delta_{i,j},\\
(H_{pq})_{i,j}
& = & \frac{1}{2}\left\{
\frac{\partial^{2}H}{\partial \bp \partial \bq} 
(\bp_{j+1},\overline{\bq}_{j+1},t_{j+1})\delta_{i,j+1} +
\frac{\partial^{2}H}{\partial \bp \partial \bq} 
(\bp_{j},\overline{\bq}_{j},t_{j})\delta_{i,j} 
\right\},\\
(H_{qq})_{i,j} 
& = & \frac{1}{4}\left\{
\frac{\partial^{2}H}{\partial \bq^{2}}
(\bp_{j+1},\overline{\bq}_{j+1},t_{j+1})
(\delta_{i,j+1} + \delta_{i,j})\right.\\
& & +
\left.\frac{\partial^{2}H}{\partial \bq^{2}}
(\bp_{j},\overline{\bq}_{j},t_{j})
(\delta_{i,j} + \delta_{i,j-1})\right\},\\
(\Delta)_{i,j}
& = & (\delta_{i,j} - \delta_{i,j+1})/\Delta t.
\label{element}
\end{eqnarray}
Therefore 
\begin{eqnarray}
Z(T) & = & 
\sum_{p.o.} \exp \left[\frac{i}{\hbar}R_{cl}\right]
\lim_{N\rightarrow\infty}\int\frac{d\bX_{N}}{(2\pi\hbar)^{Nn}}
\exp \left[\frac{i}{2\hbar}
\bX_{N}^{T}\Dslash_{N}\bX_{N}\right],
\label{spa}\\
& = &
\sum_{p.o.} \exp \left[\frac{i}{\hbar}R_{cl}\right]
\lim_{N\rightarrow\infty}
\frac{e^{i\frac{\pi}{4}(\mu_{+,N}-\mu_{-,N})}}{\sqrt{|\det\Dslash_{N}|}}
\label{sc}
\end{eqnarray}
where
\begin{eqnarray}
R_{cl}    & = & \oint(\bp_{cl}\dot{\bq}_{cl} - H(\bp_{cl},\bq_{cl},t)) dt,\\
\bX_{N} & = & (\delta\bp_{1},\delta\bp_{2},...,\delta\bp_{N},
\delta\bq_{1},\delta\bq_{2},....,\delta\bq_{N}),\\
\Dslash_{N}     & = & 
\Delta t \left(
\begin{array}{cc}
- H_{pp}                & \Delta - H_{pq} \\
\Delta^{T} - H_{pq}^{T} & - H_{qq}
\end{array}
\right), \\ 
\label{defD}
\end{eqnarray}
and $\mu_{+,N}$ ($\mu_{-,N}$) is the number of positive (negative)
integers.
Since $\Delta$ becomes differential operator in the limit 
$N\rightarrow \infty$, $Z$ can be written as
\begin{equation}
Z(T) = \sum_{p.o.} 
\frac{e^{-i\frac{\pi}{2}\mu}}{\sqrt{\left|\Det \Dslash\right|}}
\exp \left[\frac{i}{\hbar}R_{cl}\right].
\end{equation}
Here, the differential operator $\Dslash$ is defined as
\begin{equation}
\Dslash = J \frac{d}{dt} - H^{''}(\bp_{cl}(t),\bq_{cl}(t),t). 
\end{equation}
$J$ and $H^{''}$ are $2n\times 2n$ matrices defined as
\footnote{Note that the definitions of $\Dslash$ and $J$ are different
from those in the reference \cite{sugita}. (The signs are opposite.) 
Hence the definition of the Maslov index (\ref{defMaslov}) is also different
from that in \cite{sugita}. We follow the definitions in \cite{levit2}
in this paper.}
\begin{equation}
J = \left(
\begin{array}{cc}
0 & I \\
-I & 0
\end{array} \right),
\end{equation}
\begin{equation}
H^{''} = \left(
\begin{array}{cc}
\frac{\partial^{2}H}{\partial \bp^{2}} & 
\frac{\partial^{2}H}{\partial \bp\partial\bq} \\
\frac{\partial^{2}H}{\partial \bq\partial\bp} &
\frac{\partial^{2}H}{\partial \bq^{2}}
\end{array}\right).
\end{equation}

The functional determinant of $\Dslash$ is defined
as the limit of the determinant of $\Dslash_{N}$:
\begin{equation}
\Det \Dslash = \lim_{N\rightarrow\infty}\det \Dslash_{N}.
\end{equation}
$\mu$ is the Maslov index of the periodic orbit, which is defined
as
\begin{equation}
\mu = \lim_{N\rightarrow\infty}\frac{\mu_{-,N} - \mu_{+,N}}{2},
\label{defMaslov}
\end{equation}
Since $\Dslash_{N}$ is even dimensional
matrix, $\mu_{-,N} - \mu_{+,N}$ is also even and $\mu$ is integer 
if there is no zero-mode.

\subsection{Relation between Maslov index and Morse index}
Let us consider the case in which the Hamiltonian of the system 
is given by
\begin{equation}
H = \frac{\hat{\bp}^{2}}{2m} + V(\hat{\bq},t).
\end{equation}
In this case, the second variation of action around
the periodic orbit is
\begin{equation}
\frac{1}{2}\bX_{N}^{T}\Dslash_{N}\bX_{N} =
- \sum_{j=1}^{N}\left\{\frac{\Delta t}{2m}\delta\bp_{j}^{2} - 
\delta\bp_{j}(\delta\bq_{j}-\delta\bq_{j-1})\right\} -
\frac{\Delta t}{2}\sum_{i,j =1}^{N}V^{''}_{i,j}\delta\bq_{i}\delta\bq_{j},
\end{equation}
where $V^{''}_{i,j}$ is defined as
\begin{eqnarray}
V^{''}_{i,j} & = & \frac{1}{4}\left\{ 
\frac{\partial^{2}V}{\partial \bq^{2}}
(\overline{\bq}_{j+1},t_{j+1})\delta_{i,j+1} +
\frac{\partial^{2}V}{\partial \bq^{2}}
(\overline{\bq}_{j+1},t_{j+1})\delta_{i,j} 
\right. \\&& \left.+
\frac{\partial^{2}V}{\partial \bq^{2}}
(\overline{\bq}_{j},t_{j})\delta_{i,j} +
\frac{\partial^{2}V}{\partial \bq^{2}}
(\overline{\bq}_{j},t_{j})\delta_{i,j-1}
\right\}.
\end{eqnarray}
This quadratic form can be transformed as
\begin{eqnarray}
\frac{1}{2}\bX_{N}^{T}\Dslash_{N}\bX_{N} & = &
- \sum_{j=1}^{N}\frac{\Delta t}{2m}
\left\{\delta\bp_{j}-\frac{m}{\Delta t}
(\delta\bq_{j}-\delta\bq_{j-1})\right\}^{2}\\
&& + \Delta t\left\{\sum_{j=1}^{N}\frac{m}{2}
\left(\frac{\delta\bq_{j}-\delta\bq_{j-1}}{\Delta t}\right)^{2} +
\frac{1}{2}\sum_{i,j=1}^{N}V_{i,j}^{''}\delta\bq_{i}\delta\bq_{j}\right\},\\
& = &
- \sum_{j=1}^{N}\frac{\Delta t}{2m} \delta \bp_{j}^{'2}
+ \frac{\Delta t}{2} \sum_{i,j=1}D_{\bq,N, i,j}\delta\bq_{i}\delta\bq_{j}.
\end{eqnarray}
Here, $\delta\bp_{j}^{'}=\delta\bp_{j}-
m(\delta\bq_{j}-\delta\bq_{j-1})/\Delta t$, and the second term
of RHS is the second variation of
action in coordinate space path integral (\ref{lag1}).
In the limit $N\rightarrow\infty$,$D_{\bq,N}$ reduces to 
a differential operator:
\begin{eqnarray}
D_{\bq} = - m\frac{d^{2}}{dt^{2}} + V^{''}(\bq_{cl}(t),t).
\end{eqnarray}
(Here we used the partial integration $(d\bq/dt)^{2}\rightarrow
-d^{2}\bq/dt^{2}$.)
We denote the number of negative eigenvalues of  
$Nn\times Nn$ quadratic form $D_{\bq,N}$ as $\mu_{\bq,-,N}$. 
Then the Maslov index
of the orbit is
\begin{eqnarray}
\mu & = & \lim_{N\rightarrow\infty}\frac{\mu_{-,N}-\mu_{+,N}}{2},\\
& = &  \lim_{N\rightarrow\infty}
\frac{Nn - \{(Nn-\mu_{\bq,-,N})-\mu_{\bq,-,N}\}}{2},\\
& = &  \lim_{N\rightarrow\infty} \mu_{\bq,-,N}.
\end{eqnarray}
This is the number of negative eigenvalues of $D_{\bq}$ (Morse index),
which is finite if $m$ is positive.
Therefore the Maslov index is the Morse index in coordinate space path
integral. We summarize the correspondence between 
phase space path integral and coordinate path integral in Table
\ref{vs}.
\begin{table}
\begin{center}
\begin{tabular}{|c|c|c|}
\hline 
 & phase space  & coordinate space\\ 
\hline
partition function & 
$\int {\cal D}\bp{\cal D}\bq \exp
\left[\frac{i}{\hbar}\oint\bp d\bq - Hdt\right]$ &
$\int {\cal D}\bq \exp
\left[\frac{i}{\hbar}\oint Ldt\right]$\\ 
\hline
second variation & 
$\Dslash = J\frac{d}{dt} - H^{''}$& 
$D_{\bq} = - m\frac{d^{2}}{dt^{2}} + V^{''}$\\ \hline
Maslov index & $(\mu_{-}-\mu_{+})/2$ & $\mu_{-}$ (Morse index)\\ 
\hline
\end{tabular}
\end{center}
\caption{phase space v.s. coordinate space}
\label{vs}
\end{table}

%% file: quad.tex
\section{General treatment of quadratic path integrals}\label{quadra}
In section \ref{path}, we evaluate path integral by the
stationary phase approximation, 
and it is necessary to calculate quadratic path integral like (\ref{spa}). 
In this section, we study the general structure of the 
quadratic path integrals. 
First we note that
quadratic path integrals are independent of $\hbar$. 
(We can show this explicitly
by changing the variable $\bX_{N}\rightarrow\sqrt{\hbar}\bX_{N}$ 
in (\ref{spa}).)
The period $T$ can also be normalized by $t\rightarrow Tt$. 
Therefore it is enough
to consider the case in which $\hbar = T = 1$. 

\subsection{Quadratic path integrals}
Let us consider the following quadratic Hamiltonian:
\begin{equation}
\hat{H}(\hat{\bp},\hat{\bq},t) = \frac{1}{2}\hat{\bx}^{T}h(t)\hat{\bx}. 
\end{equation}
Here, $h(t)$ is a symmetric $2n\times 2n$ matrix, and 
$\hat{\bx} = (\hat{\bp},\hat{\bq})$.
 Since this Hamiltonian
is already arranged into Weyl ordering, the corresponding classical
Hamiltonian is
\begin{equation}
H(\bp,\bq,t) = \frac{1}{2}\bx^{T} h(t)\bx,
\end{equation}
where $\bx$ is a classical $2n$ dimensional vector $(\bp,\bq)$.
Our concern is to calculate partition function
\begin{equation}
Z =  \tr \hat{U}(T=1),
\end{equation}
\begin{equation}
\hat{U}(T) =
{\rm P}\exp \left[-i\int_{0}^{T}\hat{H}(t)dt\right].
\end{equation}
$Z$ can be represented by path integral as in \S \ref{path}:
\begin{eqnarray}
Z & = & \lim_{N\rightarrow\infty}\int\frac{d\bX_{N}}{(2\pi)^{Nn}}
\exp \left[\frac{i}{2}\bX_{N}^{T}\Dslash_{N}\bX_{N}\right],\\
& = & 
\lim_{N\rightarrow\infty}
\frac{e^{-i\frac{\pi}{2}\mu_{N}}}{\sqrt{\left|\det\Dslash_{N}\right|}},
\label{Zquad}
\end{eqnarray}
where
\begin{eqnarray}
\bX_{N} & = & (\bp_{1},\bp_{2},...,\bp_{N},
\bq_{1},\bq_{2},....,\bq_{N}),\\
\Dslash_{N}     & = & 
\frac{1}{N} \left(
\begin{array}{cc}
- H_{pp}                & \Delta - H_{pq} \\
\Delta^{T} - H_{pq}^{T} & - H_{qq}
\end{array}
\right), 
\end{eqnarray}
\begin{eqnarray}
(H_{pp})_{i,j,k,l} 
& = & h(t_{j})_{k,l}\delta_{i,j},\\
(H_{pq})_{i,j,k,l}
& = & \frac{1}{2}\left\{
h(t_{j+1})_{k,n+l} \delta_{i,j+1} +
h(t_{j})_{k,n+l} \delta_{i,j}
\right\},\\
(H_{qq})_{i,j,k,l} 
& = & \frac{1}{4}\left\{
h(t_{j+1})_{n+k,n+l}(\delta_{i,j+1} + \delta_{i,j}) + 
h(t_{j})_{n+k,n+l}(\delta_{i,j} + \delta_{i,j-1})
\right\},\\
\Delta_{i,j,k,l} & = & N\delta_{k,l}(\delta_{i,j}-\delta_{i,j+1}),
\end{eqnarray}
\begin{equation}
(1\le i,j \le N, 1\le k,l \le n).
\end{equation}
The quadratic path integral (\ref{spa}) appearing in the stationary phase 
approximation is the special case of (\ref{Zquad}) in which 
\begin{equation}
h(t) = TH^{''}\left(\bp_{cl}(tT),\bq_{cl}(tT)\right).
\end{equation}

In the continuum limit, we write this path integral as 
\begin{eqnarray}
Z & = & \int {\cal D}\bx \exp 
\left[\frac{i}{2}\bx^{T}\Dslash\bx\right],\\
& = & \frac{e^{-i\frac{\pi}{2}\mu}}{\sqrt{|\Det \Dslash|}}
\label{pathcont}
\end{eqnarray}
\begin{equation}
\Dslash = J\frac{d}{dt} - h(t),
\end{equation}
\begin{equation}
\bx = (\bp,\bq).
\end{equation}
Our purpose in the following subsections  is to calculate (\ref{pathcont}).
However, It is difficult to calculate (\ref{pathcont}) directly
from the definition (\ref{Zquad}). We will explain our strategy
to calculate (\ref{pathcont}) in the next subsection.

\subsection{Outline of the discussion}\label{outline}
Our strategy to calculate 
(\ref{pathcont}) is to transform the Hamiltonian matrix $h(t)$
into a simple normal form by canonical transformations. 

Let us consider a time dependent symplectic matrix $S(t)$.
This matrix satisfies
\begin{equation}
S^{T}(t)J S(t) = J,
\end{equation}
and we assume that $S(t)$ satisfies periodic boundary condition:
\begin{equation}
S(0) = S(1).
\end{equation} 
$S(t)$ defines a canonical transformation
\begin{equation}
\bx^{'}(t) = S(t)\bx(t).
\end{equation}
The Hamiltonian matrix $h(t)$ is transformed by $S(t)$ as
\begin{equation}
h \rightarrow (S^{-1})^{T}hS^{-1} - JS\frac{dS^{-1}}{dt}.
\end{equation}

We can see this transformation from a different point of view.
As it was shown in \cite{sugita}, the set 
$\{\bx(t)|0\le t\le 1,\bx(0) = \bx(1)\}$ is considered to be
a vector bundle over $S^{1}$, and $A(t) = Jh(t)$ is a connection
on this bundle. Therefore canonical transformation defined by $S(t)$ 
is considered to be a gauge transformation. Actually it is easy to
verify that $A=Jh$ is transformed by $S$ as
\begin{equation}
A \rightarrow SAS^{-1} + S\frac{dS^{-1}}{dt}.
\end{equation}
The operator $\Dslash$ can be rewritten as
\begin{equation}
\Dslash(t) = JD(t),
\end{equation}
\begin{equation}
D(t) = \frac{d}{dt} + A(t).
\end{equation}
$D$ is a covariant derivative, and parallel transport defined by
this covariant derivative is the same as the Hamiltonian equation 
of motion:
\begin{equation}
\left\{\frac{d}{dt} + A(t)\right\}\bx(t) =
\frac{d\bx}{dt}(t) + J h(t)\bx(t) = 0.
\end{equation}
We can transform given $A(t)$ into simpler form by gauge
transformations. However, the path integral (\ref{pathcont})
is not invariable to all gauge transformations. 

\begin{figure}
\epsfxsize = 0.6\textwidth
\centerline{\epsffile{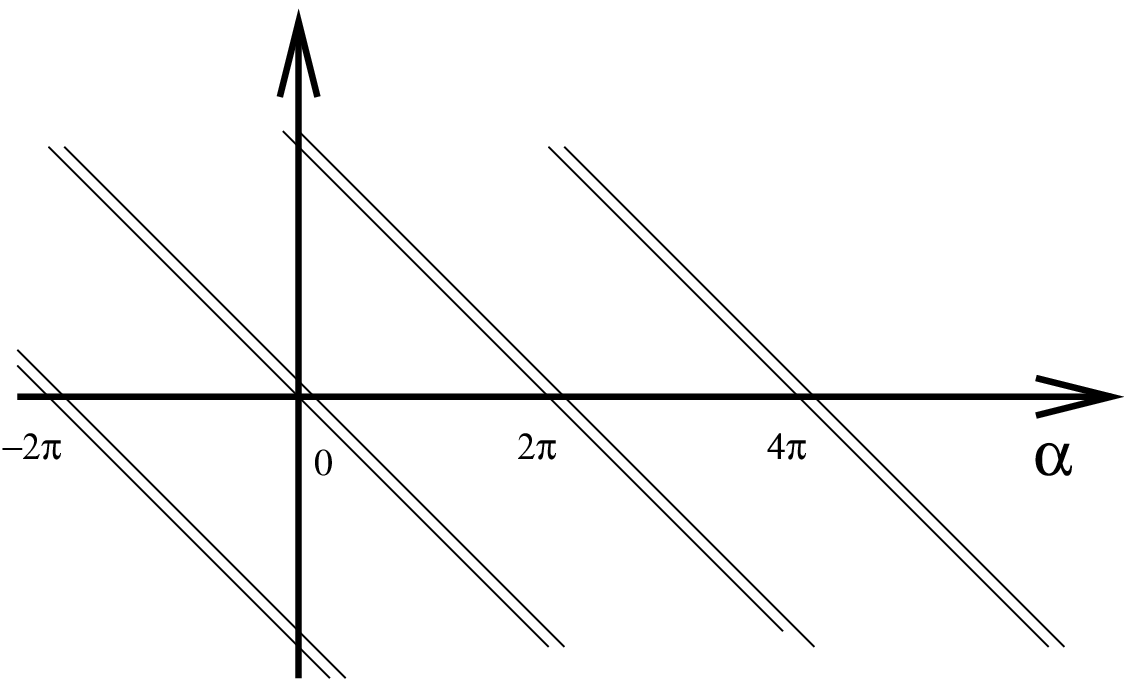}}
\caption{Flow of eigenvalues of $\Dslash$.}
\label{flow}
\end{figure}

Let us explain this point by a simple example, namely the 
time independent
one dimensional harmonic oscillator: 
\begin{equation}
H = \frac{\alpha}{2}(p^{2}+q^{2}).
\label{simple}
\end{equation}
In this case,
\begin{equation}
h = \left(
\begin{array}{cc}
\alpha & 0 \\
0 & \alpha
\end{array} \right),
\end{equation}
and
\begin{equation}
\Dslash = \left(
\begin{array}{cc}
- \alpha & d/dt \\
- d/dt & - \alpha
\end{array}\right).
\label{Dsla}
\end{equation}
Eigenvalue equation 
\begin{equation}
\Dslash \bx_{n}(t) = \epsilon_{n}\bx_{n}(t)
\end{equation} 
is easily solved as
\begin{equation}
\epsilon_{n} = -\alpha + 2n\pi,
\end{equation}
\begin{equation}
\bx_{n}(t) = \left(
\begin{array}{c}
\cos 2n\pi t \\ \sin 2n\pi t
\end{array}\right),
\left(
\begin{array}{c}
-\sin 2n\pi t \\ \cos 2n\pi t
\end{array}\right).
\end{equation}
Note that each solution is doubly degenerate. If $\alpha=0$,
$H = 0$ and the Maslov index is considered to be $0$ in this case.
As $\alpha$ changes continuously, eigenvalues $\{\epsilon_{n}\}$
also changes continuously and two of them change their sign when
$\alpha$ crosses multiples of $2\pi$ (Fig.\ref{flow}). Thus we obtain the
formula for the Maslov index:
\begin{equation}
\mu = \left\{
\begin{array}{cc}
1 + \left[\frac{\alpha}{2\pi}\right] & (\alpha\ne 2m\pi)\\
2m & (\alpha = 2m\pi)
\end{array}
\right. .
\end{equation}
(See also appendix \ref{direct}).

The canonical transformation defined by
\begin{equation}
S_{k}(t) = \left(
\begin{array}{cc}
\cos 2k\pi t & - \sin 2k\pi t \\
\sin 2k\pi t & \cos 2k\pi t
\end{array}\right)
\end{equation}
changes the Hamiltonian as
\begin{equation}
\frac{\alpha}{2}(p^{2}+q^{2}) \rightarrow
\frac{\alpha + 2k\pi}{2}(p^{2}+q^{2}).
\end{equation}
Therefore the Maslov index changes as
\begin{equation}
\mu \rightarrow \mu + 2k,
\label{2k}
\end{equation}
though the spectrum of $\Dslash$ is unchanged.

This relation means that the sign of the path integral 
changes if $k$ is odd. Let us discuss this result from 
a different point of view.

The classical time evolution operator of this system is 
represented by the matrix
\begin{equation}
V(t) = \left(
\begin{array}{cc}
\cos\alpha t & -\sin\alpha t \\
\sin\alpha t & \cos\alpha t
\end{array}
\right).
\end{equation}
On the other hand, the quantum time evolution operator 
$\hat{U}(t)=\exp (-i\hat{H}t)$ acts on coherent states as 
\begin{equation}
\hat{U}(t)|z\rangle = 
\exp (-i\alpha t/2)|\exp(-i\alpha t)z\rangle.
\end{equation}
We can see that the correspondence between
$\hat{U}(t)$ and $V(t)$ is not 1-to-1, but 2-to-1.
For example, suppose that $t=1$ and $\alpha$ changes to $\alpha + 2k\pi$.
$V(t)$ is unchanged by this transformation,
but $\hat{U}(t)$ changes to $(-1)^{k}\hat{U}(t)$.
This is consistent with the relation (\ref{2k}).

In general, the Maslov index 
changes if the gauge transformation is topologically non-trivial.
(This is considered to be a global anomaly.\cite{witten,elitzure})
The gauge transformation $S(t)$ is an element of the fundamental group
of $Sp(2n,R)$, and this group is isomorphic to $Z$ \cite{littlejohn}, 
namely,
\begin{equation}
\pi_{1}(Sp(2n,R))\simeq Z.
\end{equation} 
Therefore each gauge transformation has an integer $k$ as
a winding number, and $\mu$ is changed by this transformation
as
$\mu \rightarrow \mu + 2k$.
Hence
we should use only topologically trivial gauge transformations
when we transform $A$ into a normal form.
 
In the following subsections, we formulate canonical transformations
for quadratic path integrals rigorously, and show normal forms of
these path integrals. In \S \ref{meta}, we introduce metaplectic
group following the reference \cite{littlejohn}.
Metaplectic group is a unitary representation of symplectic
group, and the 2-to-1 correspondence between the two groups
is the origin of the anomaly as we saw in the Harmonic oscillator
model. In \S \ref{cano}, we formulate canonical transformation
using metaplectic gruop. In \S \ref{curve}, we discuss the
classical time evolution operator $V(t)$. Since this is a 
symplectic matrix with a parameter $t$, we can regard
$V$ as a curve on the group manifold of $Sp(2n,R)$.
Canonical transformations are represented by 
deformations of this curve.
In \S \ref{normalform}, we define normal forms of quadratic
path integrals, and calculate them. Details of the calculations
are shown in appendix \ref{direct}, and the results are summarized
in \S \ref{absolu}. We derive the formula for the Maslov index
of the n-repetition of a orbit in \S \ref{repeti}.

\subsection{Canonical transformations for quadratic path integrals}
\subsubsection{Metaplectic group}\label{meta}
In this section, we introduce Metaplectic group, which is
a quantum counterpart of symplectic group \cite{littlejohn}. 
Let us consider a linear canonical transformation specified by
a symplectic matrix $S\in Sp(2n,R)$:
\begin{equation}
\bx^{'} = S\bx.
\end{equation}
The metaplectic operator corresponding to $S$ is the unitary operator
$M(S)$ specified by the relation
\begin{equation}
M^{\dagger}(S)\hat{\bx}M(S) = S\hat{\bx}.
\label{Mdef}
\end{equation}
Metaplectic group $Mp(2n,R)$ is composed of such unitary operators,
and locally isomorphic to $Sp(2n,R)$.
Let us consider an infinitesimal canonical transformation
\begin{equation}
S(\epsilon) = 1 - \epsilon A.
\label{Sinfini}
\end{equation}
$A$ satisfies
\begin{equation}
A^{T}J + JA = 0,
\end{equation}
which can be written as
\begin{equation}
A = Jh,
\end{equation}
where h is a symmetric matrix.
Corresponding metaplectic operator is
\begin{equation}
M(S(\epsilon)) = 1 - \frac{i\epsilon}{2}\hat{\bx}^{T}h\hat{\bx}.
\label{Minfini}
\end{equation}
It is easy to verify that (\ref{Minfini}) and (\ref{Sinfini})
satisfy (\ref{Mdef}), using the commutation relation
\footnote{Note that we assume $\hbar = 1$ in this section.
If $\hbar\ne 1$,
$[\hat{x}_{\alpha},\hat{x}_{\beta}] = - i \hbar J_{\alpha\beta}$ and
$M(\epsilon) = I - \frac{i\epsilon}{2\hbar}\hat{\bx}^{T}h\hat{\bx}$.}
\begin{equation}
[\hat{x}_{\alpha},\hat{x}_{\beta}] = - i J_{\alpha\beta}.
\end{equation}

Let us consider a symplectic matrix $V(t)$ specified by the equation
\begin{equation}
\frac{d}{dt}V(t) = -Jh(t)V(t),
\end{equation}
\begin{equation}
V(0) = 1.
\end{equation}
$V(t)$ can be represented as
\begin{equation}
V(t) = {\rm P}\exp [-\int_{0}^{t}A(t^{'})dt^{'}],
\end{equation}
where
\begin{equation}
A(t) = Jh(t).
\end{equation}
The corresponding metaplectic operator $\hat{U}(t)$ is
uniquely determined by the following equations:
\begin{equation}
\frac{d}{dt}\hat{U}(t) = -i\hat{H}(t) \hat{U}(t),
\end{equation}
\begin{equation}
\hat{U}(0) = 1,
\end{equation}
\begin{equation}
\hat{H}(t) = \frac{1}{2}\hat{\bx}^{T}h(t)\hat{\bx}.
\end{equation}
$\hat{U}(t)$ can be represented as
\begin{equation}
\hat{U}(t) = {\rm P}\exp[-i\int_{0}^{t}\hat{H}(t^{'})dt^{'}].
\end{equation}
$V(t)$ is the classical time evolution operator for
the quadratic Hamiltonian $H(t) = \bx^{T}h(t)\bx$, and 
$\hat{U}(t)$ is the corresponding quantum time evolution
operator. $V(1)$ (classical time evolution operator for the period)
is called the monodromy matrix.
We summarize the correspondence between classical
and quantum systems in Table \ref{cq}.
\begin{table}
\begin{center}
\begin{tabular}{|c|c|c|c|}
\hline 
 & Hamiltonian & eq. of motion & time evolution operator \\
\hline 
classical & $H(t) = \frac{1}{2}\bx^{T}h(t)\bx$ &
$J\frac{d}{dt}\bx = h(t)\bx$ & 
$V(t) = {\rm P}\exp [-\int_{0}^{t}A(t^{'})dt^{'}]$\\
 & & (Hamilton eq.) & 
$A(t) = Jh(t)$,$V(t)\in Sp(2n,R)$ \\
\hline 
quantum & $\hat{H}(t) = \frac{1}{2}\hat{\bx}^{T}h(t)\hat{\bx}$ &
$i\frac{d}{dt}|\varphi\rangle = \hat{H}(t)|\varphi\rangle$ & 
$\hat{U}(t) = {\rm P}\exp[-i\int_{0}^{t}\hat{H}(t^{'})dt^{'}]$ \\ 
 & & (Shr\"{o}dinger eq.) & 
$\hat{U}(t)\in Mp(2n,R)$\\
\hline
\end{tabular}
\end{center}
\caption{Correspondence between classical and quantum systems for
quadratic Hamiltonian.}
\label{cq}
\end{table}

Although $Mp(2n,R)$ and $Sp(2n,R)$ are isomorphic locally, they
are not isomorphic globally; there is a 2-to-1 correspondence 
between them. Therefore if a closed curve on symplectic group 
$S(t) (S(0) = S(1))$ has winding number $k$, the metaplectic
operators corresponding to $S(0)$ and $S(1)$ are not necessarily
the same, but 
\begin{equation}
M(S(1)) = (-1)^{k}M(S(0)).
\end{equation}
This relation is the origin of the anomaly.
Note that the relation between $Sp(2n,R)$ and $Mp(2n,R)$ is very similar to 
the well-known relation between $SO(3)$ and $SU(2)$. 
The point is that both $Sp(2n,R)$ and $SO(3)$
are not simply connected and hence allow 
two-valued representations.

\subsubsection{Canonical transformation}\label{cano}

Let us consider canonical transformations of quadratic path integrals 
using Metaplectic
operators. In continuum notation, a partition function
\begin{equation}
Z = \int {\cal D}\bx \exp 
\left[\frac{i}{2}\int_{0}^{1}
\bx^{T}(t)J\left(\frac{d}{dt}+A(t)\right)\bx (t) dt\right]
\label{Zcont}
\end{equation}
seems invariant under the canonical (gauge) transformation by 
$\{S(t)\}\in \pi_{1}(Sp(2n,R))$:
\begin{eqnarray}
\bx(t) & \rightarrow & S(t)\bx(t),\\
A(t) & \rightarrow & S(t)A(t)S^{-1}(t) + S(t)\frac{dS^{-1}}{dt}(t), 
\end{eqnarray}
\begin{equation}
S(0) = S(1).
\end{equation}
However, this is not correct. We must go back
to the definition of the path integral to make the discussion rigorous.

$Z$ is the trace of the time evolution operator $\hat{U}(t=1)$, which
is represented as a product of infinitesimal time evolution operators:
\begin{equation}
\hat{U}(t=1) = \prod_{j=1}^{N} \left(1-\frac{i}{N}\hat{H}(t_{j})\right)
+ O\left(\frac{1}{N}\right),
\label{prod}
\end{equation}
where $t_{j}=1/N$.
We obtain normal path integral by inserting the complete set
\begin{equation}
1 = \int d\bq_{j} |\bq_{j}\rangle\langle\bq_{j}|
\label{comp}
\end{equation} 
to (\ref{prod}):
\begin{equation}
Z = \int \prod_{j=1}^{N} d\bq_{j} \langle \bq_{j} |1 - \frac{i}{N}\hat{H}(t_{j}) |\bq_{j-1}\rangle + O\left(\frac{1}{N}\right).
\label{Z1}
\end{equation}
To formulate the canonical transformation corresponding to $S(t)$, 
we insert
\begin{equation}
1 = \int d\bq_{j}\;
M(S(t_{j}))^{\dagger}|\bq_{j}\rangle\langle\bq_{j}| 
M(S(t_{j}))
\end{equation}
instead of (\ref{comp}).
Then (\ref{Z1}) is modified as
\begin{equation}
Z = \int \prod_{j=1}^{N} d\bq_{j} 
\langle \bq_{j} |M_{j}
\left\{1 - \frac{i}{N}\hat{H}(t_{j})\right\}M_{j-1}^{\dagger}
|\bq_{j-1}\rangle 
+ O\left(\frac{1}{N}\right),
\end{equation}
where $M_{j}$ denotes $M(S(t_{j}))$.

$S(t_{j-1})$ can be expanded as
\begin{equation}
S(t_{j-1}) = 
\left\{1-\frac{1}{N}\frac{dS}{dt}(t_{j}) S(t_{j})^{-1}\right\}S(t_{j})
+ O\left(\frac{1}{N^{2}}\right).
\end{equation}
Therefore
\begin{equation}
M_{j}M_{j-1}^{\dagger} = 1 - \frac{i}{2N}
\left\{JS(t_{j})\frac{dS^{-1}}{dt}(t_{j})\right\}_{\alpha\beta}
\hat{x}_{\alpha}\hat{x}_{\beta}
+ O\left(\frac{1}{N^{2}}\right), 
\end{equation}
\begin{eqnarray}
M_{j}\hat{H}(t_{j})M_{j-1}^{\dagger} & = &
M_{j}\hat{H}(t_{j})M_{j}^{\dagger} + O\left(\frac{1}{N}\right),\\
& = & 
\frac{1}{2}h_{\alpha\beta}(t_{j})
M_{j}\hat{x}_{\alpha}M_{j}^{\dagger}M_{j}\hat{x}_{\beta}M_{j}^{\dagger}
+ O\left(\frac{1}{N}\right),\\
& = &
\frac{1}{2}h_{\alpha\beta}(t_{j})
S^{-1}_{\alpha\gamma}(t_{j})S^{-1}_{\beta\delta}(t_{j})
\hat{x}_{\gamma}\hat{x}_{\delta}
+ O\left(\frac{1}{N}\right).
\end{eqnarray}
Here we used $M(S)\hat{\bx}M(S)^{\dagger}=S^{-1}\hat{\bx}$ and
$\frac{dS}{dt}S^{-1} = S\frac{dS^{-1}}{dt}$.
Hence, for $2\le j\le N$,
\begin{equation}
\langle \bq_{j} |M_{j}
\left\{1 - \frac{i}{N}\hat{H}(t_{j})\right\}M_{j-1}^{\dagger}
|\bq_{j-1}\rangle =
\langle \bq_{j} | 1 - \frac{i}{N}\hat{K}(t_{j}) |\bq_{j-1}\rangle
+ O\left(\frac{1}{N^{2}}\right), 
\end{equation}
\begin{equation}
\hat{K}(t) = \hat{\bx}^{T}k(t)\hat{\bx},
\end{equation}
\begin{equation}
k(t) = (S(t)^{-1})^{T}h(t)S(t)^{-1}
- JS(t)\frac{dS^{-1}}{dt}(t).
\end{equation}
This is the same as the classical canonical transformation.
\footnote{Note that this simple correspondence between
classical and quantum canonical transformation is due to 
Weyl ordering of the operators. If the Hamiltonian is written
in other order, additional terms appear and the simple 
correspondence is lost. See, for example, chapter 3 of the
reference \cite{kashiwa}.}
However, for $j=1$,
\begin{equation}
M_{1}M_{N}^{\dagger} = (-1)^{k} + O\left(\frac{1}{N}\right),
\end{equation}
where $k$ is the winding number of the closed path $\{S(t)\}$.
Therefore $Z$ is represented as
\begin{eqnarray}
Z & = & \int \prod_{j=1}^{N} d\bq_{j} 
\langle \bq_{j} |
1 - \frac{i}{N}\hat{H}(t_{j})|\bq_{j-1}\rangle 
+ O\left(\frac{1}{N}\right),\\
& = & (-1)^{k}\int \prod_{j=1}^{N} d\bq_{j} 
\langle \bq_{j} |
1 - \frac{i}{N}\hat{K}(t_{j})|\bq_{j-1}\rangle 
+ O\left(\frac{1}{N}\right),
\end{eqnarray}
and the path integral (\ref{Zcont}) is transformed as
a functional of $A$ like
\begin{equation}
Z[A^{S}(t)] = (-1)^{k}Z[A(t)],
\label{even_odd} 
\end{equation}
where $A^{S}$ is the gauge field transformed by $S$.
(\ref{even_odd}) means that the Maslov index is transformed as
\begin{equation}
\mu^{'} \equiv \mu + 2k \;\;\; ({\rm mod} 4),
\end{equation}
where $\mu^{'}$ is the Maslov index corresponding to $A^{S}(t)$.
We show the stronger result
\begin{equation}
\mu^{'} = \mu + 2k,
\label{mu_dash}
\end{equation}
in the following subsection.

\subsection{Curves on $Sp(2n,R)$}\label{curve}
To investigate quadratic path integrals in detail, 
let us examine the classical time evolution operator
$V$, which is considered to be a curve on the
group manifold of $Sp(2n,R)$. $\{V(t)\}$ is the
curve which starts from the origin, and its
end point is the monodromy matrix.
$V$ is defined by the equations
\begin{equation}
\Dslash V(t) = J\left(\frac{d}{dt} + A(t)\right)V(t) = 0,
\label{DV}
\end{equation}
\begin{equation}
V(0) = 1.
\label{initial}
\end{equation}
Note that the correspondence between $V$ and $\Dslash$ is
1-to-1. $V$ is determined from $\Dslash$ by (\ref{DV}) 
and (\ref{initial}), and $\Dslash$ is determined from $V$
by the relation
\begin{equation}
A(t) = - \frac{dV}{dt}(t) V(t)^{-1}.
\end{equation}
See also Table \ref{HAV}.
\begin{table}
\begin{center}
\begin{tabular}{|c|c|c|}
\hline
            & definition & gauge transformation\\ \hline
Hamiltonian &$H(\bp,\bq,t) = \bx^{T}h(t)\bx $& 
$(S(t)^{-1})^{T}h(t)S(t)^{-1} - JS(t)\frac{dS^{-1}}{dt}(t)$ \\
            & $h$: symmetric matrix& \\ \hline
gauge field & $A(t) = J h(t)$ & 
$S(t)A(t)S(t)^{-1} + S(t)\frac{dS^{-1}}{dt}(t)$ \\
            & $A^{T}J + JA = 0$ & \\ \hline
curve on $Sp(2n,R)$ & $(d/dt + A(t))V(t) = 0$, $V(0)=1$ & 
$S(t)V(t)S(0)^{-1}$ \\
                    & $V(t)^{T}JV(t) = J$     &  \\
\hline
\end{tabular}
\end{center}
\caption{Hamiltonian $H$, gauge field $A$ and curve $V$.
Those three have the same informations. 
$S(t)$ belongs to $Sp(2n,R)$ and satisfies $S(1)=S(0)$.
$V$ and $S$ is continuous, but we don't require smoothness of 
$V$ and $S$ in this paper. 
Therefore $H$ and $A$ allow discontinuities.}
\label{HAV}
\end{table}

\subsubsection{Gauge transformation of $V$}
Let us consider how this curve is transformed by
gauge transformations. 
If the gauge field is transformed as 
\begin{equation}
A(t) \rightarrow S(t)A(t)S(t)^{-1} + S(t)\frac{dS^{-1}}{dt}(t),
\end{equation}
solutions of (\ref{DV}) are $S(t)V(t)\times{\rm constant}$.
The constant should be chosen to be $S(0)^{-1}$ to satisfy 
(\ref{initial}). Thus $V$ is transformed as
\begin{equation}
V(t) \rightarrow S(t)V(t)S(0)^{-1}.
\end{equation}

The gauge transformation group $G$ is considered to be a set
of closed curves on $Sp(2n,R)$:
\begin{equation}
G = \{S| S:[0,1]\rightarrow Sp(2n,R),S(0)=S(1)\}.
\end{equation}
$G$ can be decomposed into two parts,
\begin{equation}
G = L\times N,
\label{decomp}
\end{equation}
where $L$ and $N$ are subgroups of $G$ defined as
\begin{equation}
L = \{l|l\in G, l(0)=1\},
\end{equation} 
\begin{equation}
N = \{n| n: \mbox{time independent element in $G$}\}
\end{equation}
$N$ is isomorphic to $Sp(2n,R)$.
(\ref{decomp}) means any $S\in G$ can be written as
\begin{equation}
S(t) = l(t)n,
\end{equation}
where $l\in L$ and $n\in N$. $l$ and $n$ are explicitly written as
\begin{eqnarray}
n & = & S(0),\\
l(t) & = & S(t)S(0)^{-1}.
\end{eqnarray}
The gauge transformation by $l\in L$ is written as
\begin{equation}
V(t) \rightarrow l(t)V(t).
\end{equation}
The monodromy matrix $M (= V(1))$ is unchanged by this transformation. 
The transformation by $n\in N$ is
\begin{equation}
V(t) \rightarrow nV(t)n^{-1},
\end{equation}
and $M$ is transformed as $nMn^{-1}$.

The elements in $G$ and $L$ are classified by winding numbers.
We refer the subset of $G$ ($L$) which has winding number $k$
as $G_{k}$ ($L_{k}$). $G_{0}$ ($L_{0}$) is a subgroup of $G$ ($L$),
and $N$ is a subgroup of $G_{0}$. 
It is obvious that
\begin{equation}
G_{0} = L_{0}\times N.
\end{equation}

It is worth pointing out, in passing, that the time translation
$t\rightarrow t+t_{0}$ is also regarded as a gauge transformation.
This transformation changes $V$ as 
\begin{equation}
V(t)\rightarrow \left\{
\begin{array}{cc}
V(t+t_{0})V(t_{0})^{-1} & (0\le t \le 1-t_{0}) \\
V(t+t_{0}-1)V(t_{0})^{-1} & (1-t_{0}\le t \le 1)
\end{array}
\right.
\end{equation}
This is a gauge transformation given by
\begin{equation}
S(t) = \left\{ \begin{array}{cc}
V(t+t_{0})V(t)^{-1} & (0\le t \le 1-t_{0}) \\
V(t+t_{0}-1)V(t)^{-1} & (1-t_{0}\le t \le 1)
\end{array}\right.
\end{equation} 
which can be shrunk to a point continuously as $t_{0}\rightarrow 0$.
Therefore the time translation is regarded as a gauge transformation
in $G_{0}$.

\subsubsection{Continuous deformation of $V$ and its Maslov index}
\label{defV}

Let us consider deformations of $V$.
If the deformation is a gauge transformation, 
the monodromy matrix $M$ is transformed as $nMn^{-1}$ by
a symplectic matrix $n$. The converse also holds. 
Let $V_{1}$ and $V_{2}$ be the curves which start from the 
origin. If there is a symplectic matrix $n$ which satisfies
$V_{1}(1) = nV_{2}(1)n^{-1}$,
the matrix $S(t)$ defined by
\begin{equation}
S(t) = V_{1}(t)nV_{2}(t)^{-1}
\end{equation}
satisfies $S(0) = S(1) = n$ and 
$V_{1}(t) = S(t)V_{2}(t)S(0)^{-1}$. 
Therefore $V_{1}$ and $V_{2}$ are connected by the gauge 
transformation $S$. Thus we conclude that two curves are connected
by a gauge transformation if and only if the end-points of 
two curves are in the same conjugacy class. As a special case,
if two curves have the same end-point (monodromy matrix), there
is a gauge transformation which connects two curves. In this case,
$n=1$ and $S(t)=V_{1}(t)V_{2}(t)^{-1}$ is an element of $L$.

\begin{figure}
\epsfxsize = 0.6\textwidth
\centerline{\epsffile{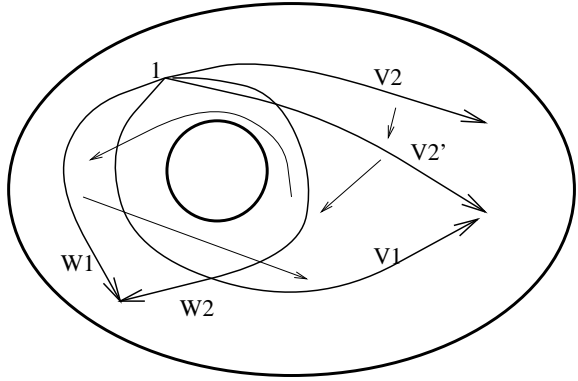}}
\caption{Deformations of curves on $Sp(2n,R)$.}
\label{Vtrans}
\end{figure}

Since there is a 1-to-1 correspondence between $\Dslash$ and
$V$, $\Dslash$ and its spectrum changes continuously
as we deform $V$ continuously. We can see relative changes of Maslov
index by analyzing the points where the signs of diagonal elements
of $\Dslash$ changes. \footnote{Strictly speaking, we should determine
the Maslov index from discretized matrix $\Dslash_{N}$. However, 
when we deform $V$ (or $A$) continuously, 
relative changes of the Maslov index are dominated by the 
low-frequency part of $\Dslash_{N}$, which is approximated
well by $\Dslash$ if $N$ is large enough.
Therefore we can determine
the relative changes of the Maslov index from $\Dslash$.}
$\Dslash$ can be diagonalized explicitly, and what is important is that
the diagonal elements are unchanged by gauge transformations. 
(See appendix \ref{diag}.) 
Let $V_{1}$ and $V_{2}$ be curves which are connected by a 
gauge transformation $S$. If $S$ belongs to $G_{0}$, 
$S$ can be shrunk to 1 continuously. Therefore $V_{1}$ and $V_{2}$
can also be united continuously, and diagonal elements of 
$\Dslash$ is unchanged throughout this process. 
Thus we conclude that if two curves are connected by $G_{0}$,
two curves have the same Maslov index. However, If two curves
are connected by $G_{k}\; (k\ne 0)$, two curves have different
Maslov indices. Now we can prove the following important statement:\\
If two curves $V_{1}$ and $V_{2}$ satisfy the relation
\begin{equation}
V_{1}(t) = S(t)V_{2}(t)S(0)^{-1},
\label{assump1}
\end{equation}
\begin{equation}
S \in G_{k},
\label{assump2}
\end{equation} 
the corresponding Maslov indices $\mu_{1}$ and $\mu_{2}$ 
satisfy
\begin{equation}
\mu_{2} = \mu_{1} + 2k.
\label{conclusion}
\end{equation}
[proof]\\
$V_{2}$ can be deformed to $V_{2}^{'}$, 
which is defined as
\begin{equation}
V_{2}^{'}(t) = S(0)V_{2}(t)S(0)^{-1}.
\end{equation} 
Since the constant matrix $S(0)$ belongs to $G_{0}$, 
$V_{2}$ and $V_{2}^{'}$ have the same Maslov index.
$V_{1}$ and $V_{2}^{'}$ have the same end-point.
(See Fig. \ref{Vtrans}.)
We already know that (\ref{conclusion}) holds for
some cases. For example, let us consider the Hamiltonian
\begin{equation}
H(\bp,\bq,\alpha) = \frac{\alpha}{2}(p_{1}^{2}+q_{1}^{2}).
\end{equation}
Let $W_{1}$ and $W_{2}$ be the curves corresponding to
$H(\bp,\bq,\alpha)$ and $H(\bp,\bq,\alpha + 2k\pi)$.
$W_{1}$ and $W_{2}$ have the same end-point, and
the difference between the Maslov indices of $W_{1}$
and $W_{2}$ is 2k. (See section \ref{outline} and 
appendix \ref{direct}.)
Then deform $V_{1}$ continuously to $V_{2}^{'}$ through
$W_{1}$ and $W_{2}$ (Fig. \ref{Vtrans}). The difference of 
the Maslov index between $V_{1}$ and $V_{2}^{'}$ is the same
as that of $W_{1}$ and $W_{2}$, because the change of the
Maslov index during the deformation $V_{1}\rightarrow W_{2}$
is canceled by that of $W_{2}\rightarrow V_{2}^{'}$ if the end-point
go through the same path during the two processes. Therefore
the difference of the Maslov index between $V_{1}$ and $V_{2}$
is also $2k$. (Q.E.D.)

(\ref{conclusion}) means that two curves considered to be equivalent
if two curves are connected by $G_{0}$ (not $G$). We classify 
quadratic Hamiltonians by this equivalence relation in the
next subsection.

\subsubsection{Universal covering of $Sp(2n,R)$
and generalized monodromy matrices}

Let us consider the set of curves on $Sp(2n,R)$ which starts
from the origin. We refer this set as $F$:
\begin{equation}
F = \{V|V: [0,1]\rightarrow Sp(2n,R), V(0) = 1\},
\end{equation}
which is isomorphic
to the set of quadratic Hamiltonians and the set of gauge fields.
Two elements $V_{1}$ and $V_{2}$ in $F$ is considered to be equivalent
if they are connected by a gauge transformation which belongs to 
$G_{0}$. We express this relation as
\begin{equation}
V_{1} \Gzerosim V_{2}.
\end{equation}
If $V_{1} \Gzerosim V_{2}$, $V_{1}$ and $V_{2}$ 
have the same Maslov index.
Therefore the Maslov index is determined uniquely on the
quotient set $F/\Gzerosim$.

\begin{figure}
\epsfxsize = 0.8\textwidth
\centerline{\epsffile{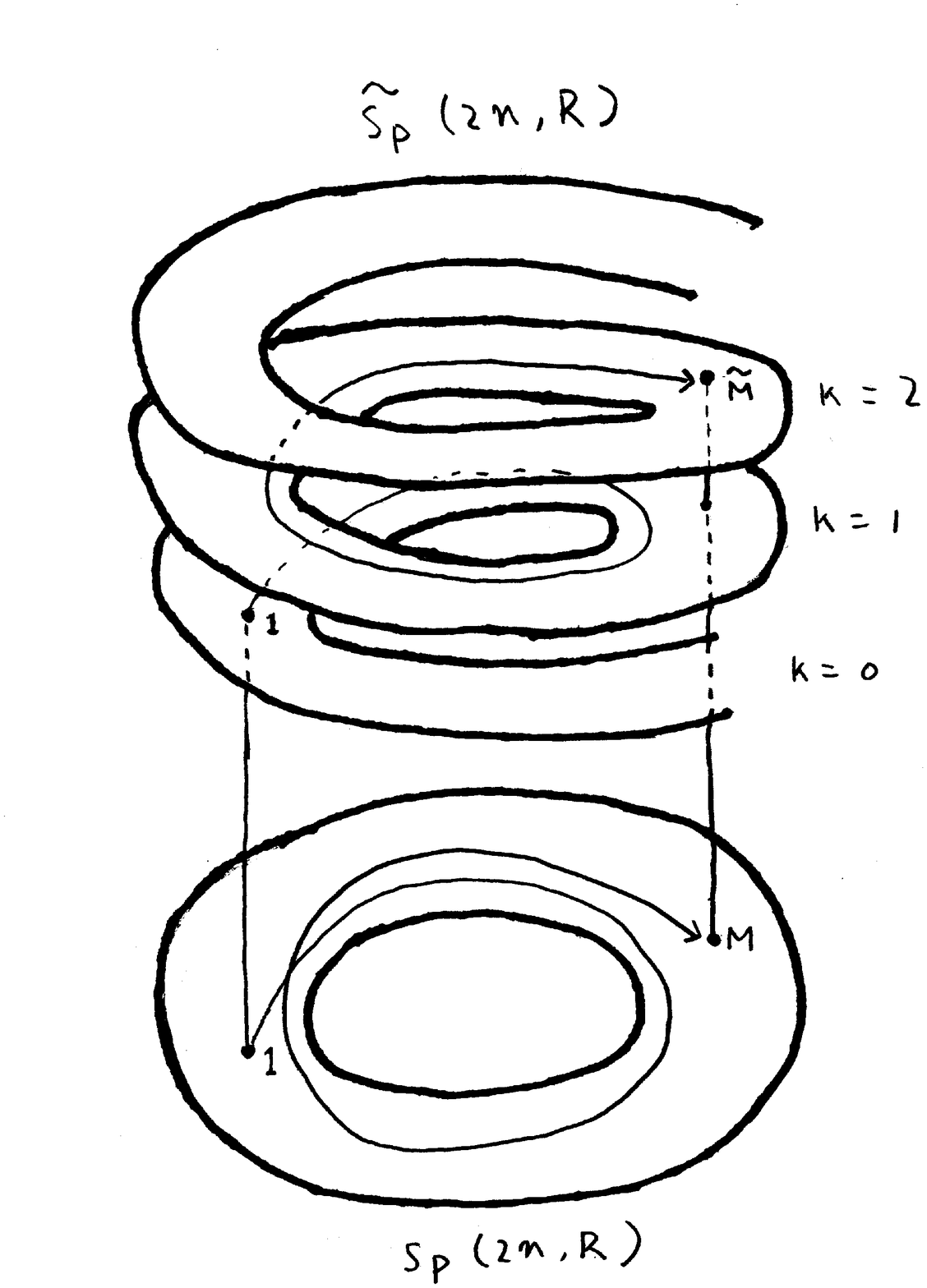}}
\caption{$Sp(2n,R)$ and $\tilde{Sp}(2n,R)$.}
\label{covering}
\end{figure}

Before turning to the division by $G_{0}$, let
us discuss the division by $L_{0}$, which is
a subgroup of $G_{0}$. If $V_{1}$ and $V_{2}$ are connected by
an element in $L_{0}$ (i.e. $V_{1}\Lzerosim V_{2}$), two curves
have the same terminal points and the closed curve made by these
two curves can be shrunk to a point continuously. The converse
also holds. (It is obvious that $\{V_{1}(t)V_{2}(t)^{-1}\}$ belongs
to $L_{0}$.) Therefore the quotient set $F/\Lzerosim$ is the
set of homotopy classes, which is simply connected 
by definition. Therefore this is the universal covering space
of $Sp(2n,R)$:
\begin{equation}
F/\Lzerosim = \tilde{Sp}(2n,R).
\end{equation}
An element in $\tilde{Sp}(2n,R)$ is specified by a symplectic
matrix (monodromy matrix $M$) and an integer (winding number $k$).
(See Fig. \ref{covering})
This pair $\tilde{M} = (M,k)$ is regarded as
a generalized monodromy matrix. We give a definition of
the winding number in \S \ref{G0}.

Let us discuss the group structure of 
$F$ and $\tilde{Sp}(2n,R)$.
We can define the product of two elements $V$ and $W$ in F as
\begin{equation}
(VW)(t) = V(t)W(t) \;\;\; (0\le t\le 1).
\end{equation}
It is obvious that $F$ becomes a group by this product.
Therefore we define the product in 
$\tilde{Sp}(2n,R) = F/\Lzerosim$ as
\begin{equation}
[V][W] = [VW],
\end{equation}
where [V],[W] and [VW] denote the homotopy classes of
V,W and VW.
It is easy to verify that this product is well-defined,
and $\tilde{Sp}(2n,R)$ also becomes a group by this product.

We can define another product in $F$ as
\begin{equation}
(V*W)(t) = \left\{
\begin{array}{cc}
W(2t) & (0\le t < 1/2), \\
V(2t-1)W(1) & (1/2\le t\le 1).
\end{array}\right.
\end{equation}
Note that this product doesn't satisfy the associative law, i.e. 
$V*(W*X)\ne (V*W)*X$.
However, $VW$ and $V*W$ is homotopic. We can define a map
$f:[0,1]\times[0,1]\rightarrow Sp(2n,R)$ which satisfies
$f(t,0)=(VW)(t)$ and $f(t,1)=(V*W)(t)$ as
\begin{equation}
f(t,s) = V(t,s)W(t,s)\;\;\; (0\le t \le 1, 0\le s\le 1),
\end{equation}
\begin{equation}
V(t,s) = \left\{
\begin{array}{cc}
1 & (0\le t< s/2), \\
V\left(\frac{2t-s}{2-s}\right)  & (s/2 \le t \le 1).
\end{array}
\right.
\end{equation}
\begin{equation}
W(t,s) = \left\{
\begin{array}{cc}
W\left(\frac{2}{2-s}t\right) & (0\le t< 1 - s/2),\\
W(1) &         (1 - s/2\le t \le 1), 
\end{array}
\right.
\end{equation} 
Therefore, in $\tilde{Sp}(2n,R)$,
\begin{equation}
[V][W] = [VW] = [V*W].
\end{equation}

\subsection{Normal forms of quadratic Hamiltonians}\label{normalform}
It was observed in the preceding subsection that 
\begin{eqnarray}
F/\Lsim & = & Sp(2n,R), \\
F/\Lzerosim & = & \tilde{Sp}(2n,R).
\end{eqnarray}
We are now ready to consider $F/\Gsim$ and $F/\Gzerosim$.
Since $G = L\times N$, $F/\Gsim$ is the set of conjugacy classes
of $Sp(2n,R)$. Therefore an element in $F/\Gzerosim$ is specified
by a conjugacy class of $Sp(2n,R)$ and a winding number.
In the following, we show representatives of these quotient sets.

In this subsection, we change the definition of the 
$2n\times 2n$ matrix $J$ as
\begin{equation}
J = \left(\begin{array}{ccccc}
j & & & & \\
& j & & & \\
& & . & & \\
& & & . & \\
& & & & j
\end{array}
\right),
\end{equation}
\begin{equation}
j = \left(\begin{array}{cc}
0 & 1\\
-1 & 0
\end{array}\right),
\end{equation}
for convenience.

\subsubsection{Normal forms of monodromy matrices}\label{monod}
First we consider representatives of the quotient set
\begin{equation}
F/\Gsim = \{\mbox{conjugacy class of } Sp(2n,R)\}.
\end{equation}
Our task here is to transform
a given monodromy matrix $M\in Sp(2n,R)$ into a simple normal
form $SMS^{-1}$, where $S$ is also a symplectic matrix.
In other words, we choose the special
symplectic basis in which $M$ has a simple form.

\begin{figure}
\epsfxsize = 0.8\textwidth
\centerline{\epsffile{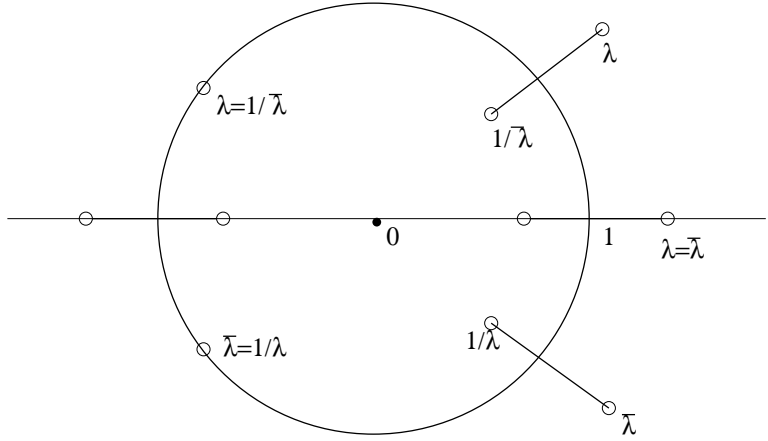}}
\caption{Eigenvalues of a symplectic matrix in the complex plane}
\label{eigen}
\end{figure}

Let us consider the eigenvalue equation
\begin{equation}
M\bx = \lambda\bx.
\end{equation}
First we assume that $M$ has no degenerate eigenvalue.
If $\lambda$ is an eigenvalue of $M$,
$\lambda^{*}$ and $1/\lambda$ are also eigenvalues of $M$
\cite{arnold}. (Fig.\ref{eigen})

Therefore we can classify eigenvalues of 
$M$ into four types, and $M$ become a block diagonal matrix
in the basis made of the eigen vectors. (If some of the eigenvectors
are complex, we use the real parts and imaginary parts of them as
the basis.)
\begin{equation}
M = \left(
\begin{array}{ccccc}
m_{1} & & & & \\
& m_{2} & & & \\
& & .     & & \\
& & & .     &  \\
& & & & m_{l}
\end{array} \right).
\end{equation}

In the following, we list up the four types of eigenvalues
and the normal forms of the monodromy matrix. 
\begin{enumerate}
\item elliptic type: $\lambda = e^{\pm i\alpha}$. ($0<\alpha < 2\pi$) 
\begin{equation}
m = \left(
\begin{array}{cc}
\cos\alpha & - \sin\alpha \\
\sin\alpha & \cos\alpha
\end{array}
\right).
\end{equation}
\item hyperbolic type: $\lambda = e^{\pm \beta}$. 
\begin{equation}
m = \left(
\begin{array}{cc}
e^{\beta} & 0 \\
0 & e^{-\beta} 
\end{array}
\right).
\end{equation}
\item inverse hyperbolic type: $\lambda = - e^{\pm\beta}$.
\begin{equation}
m = \left(
\begin{array}{cc}
- e^{\beta} & 0 \\
0 & - e^{-\beta}
\end{array}
\right).
\end{equation}
\item loxodromic type: $\lambda = e^{\pm i\alpha \pm \beta}$.
($0 < \alpha < \pi$)
\footnote{In the elliptic case, the normal forms corresponding to $\alpha$ 
and $2\pi - \alpha $ are not conjugate, though two matrices have the
same eigenvalues. However, in the loxodromic case, two matrices
corresponding to $\alpha$ and $2\pi-\alpha$ are conjugate. That's why
we choose $\alpha$ as $0<\alpha < \pi$ in this case. See 
Appendix \ref{elliptic} and the
reference \cite{arnold}, \S 42.} 
\begin{equation}
m = \left(
\begin{array}{cccc}
e^{\beta}\cos\alpha & 0 & e^{\beta}\sin\alpha & 0 \\
0 & e^{-\beta}\cos\alpha & 0 & e^{-\beta}\sin\alpha \\
-e^{\beta}\sin\alpha & 0 & e^{\beta}\cos\alpha & 0 \\
0 & -e^{-\beta}\sin\alpha & 0 & e^{-\beta}\cos\alpha
\end{array}
\right).
\end{equation}
\end{enumerate}
Here, $\alpha$ and $\beta$ are real numbers.

If $M$ have degenerate eigenvalues, we must treat such cases
separately. In this paper, we treat only the most important case,
parabolic type. In this case, the eigenvalue 1 is doubly degenerate
and we can choose the basis $\bx_{\alpha}\; (\alpha = 1,2)$ as
\footnote{$\gamma$ can be transformed to $\pm 1$. 
($\bx_{1}\rightarrow\sqrt{|\gamma|}\bx_{1}$,
 $\bx_{2}\rightarrow(1/\sqrt{|\gamma|})\bx_{2}$.)
However, we leave $\gamma$ as a parameter for convenience.}
\begin{eqnarray}
M\bx_{1} & = & \bx_{1}, \\
M\bx_{2} & = & \bx_{2} - \gamma\bx_{1}.
\end{eqnarray}
The monodromy matrix of this part is 
\begin{equation}
m = \left(
\begin{array}{cc}
1 & -\gamma \\
0 & 1
\end{array}
\right).
\end{equation}

\subsubsection{Normal forms of Hamiltonians}\label{G0}

Now we can choose representatives of $F/\Gsim$ and $F/\Gzerosim$.
Normal forms of Hamiltonians corresponding to them are also determined.

First we show
the representatives of $F/\Gsim$ and the normal form of the Hamiltonian
corresponding to the types of of eigenvalues.
\begin{itemize}
\item elliptic type
\begin{equation}
V(t) = \left(
\begin{array}{cc}
\cos\alpha t & - \sin\alpha t \\
\sin\alpha t & \cos\alpha t
\end{array}
\right).
\end{equation}
\begin{equation}
H = \frac{\alpha}{2}(p^{2}+q^{2}).
\end{equation}
\item hyperbolic type
\begin{equation}
V(t) = \left(
\begin{array}{cc}
e^{\beta t} & 0 \\
0 & e^{-\beta t} 
\end{array}
\right).
\end{equation}
\begin{equation}
H = - \beta pq.
\end{equation}
\item inverse hyperbolic type
\begin{equation}
V(t) = \left\{
\begin{array}{cc}
\left(
\begin{array}{cc}
\cos 2\pi t & - \sin 2\pi t \\
\sin 2\pi t & \cos 2\pi t
\end{array} \right)
& (0\le t < 1/2), \\
\left(
\begin{array}{cc}
- e^{\beta(2t-1)} & 0 \\
0 & - e^{-\beta(2t-1)}
\end{array}
\right) & (1/2\le t \le 1)
\end{array}\right.
\end{equation}
\begin{equation}
H = \left\{
\begin{array}{cc}
\pi (p^{2}+q^{2}) & (0\le t < 1/2), \\
- 2\beta pq & (1/2 \le t \le 1).
\end{array}
\right. 
\end{equation}
\item loxodromic type
\begin{equation}
V(t) = \left(
\begin{array}{cccc}
e^{\beta t}\cos\alpha t & 0 & e^{\beta t}\sin\alpha t & 0 \\
0 & e^{-\beta t}\cos\alpha t & 0 & e^{-\beta t}\sin\alpha t \\
-e^{\beta t}\sin\alpha t & 0 & e^{\beta t}\cos\alpha t & 0 \\
0 & -e^{-\beta t}\sin\alpha t & 0 & e^{-\beta t}\cos\alpha t
\end{array}
\right).
\end{equation}
\begin{equation}
H = \alpha (p_{1}q_{2} - p_{2}q_{1}) 
   - \beta (p_{1}q_{1} + p_{2}q_{2}).
\end{equation}
\item parabolic type
\begin{equation}
V(t) = \left(
\begin{array}{cc}
1 & -\gamma t \\
0 & 1
\end{array}
\right)
\end{equation}
\begin{equation}
H = \frac{\gamma}{2} q^{2}.
\end{equation}
\end{itemize} 
 
\begin{figure}
\epsfxsize = 0.6\textwidth
\centerline{\epsffile{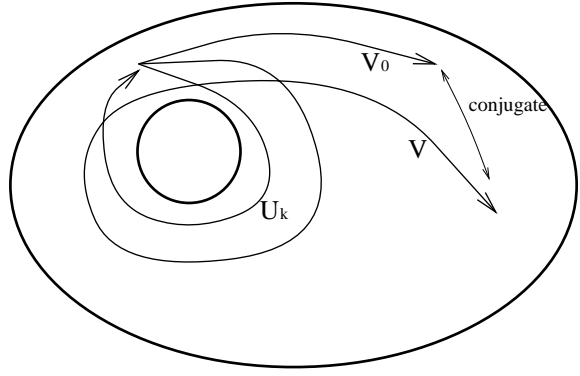}}
\caption{Normal form of $V$}
\label{normal}
\end{figure}

Let us consider a given time-dependent quadratic Hamiltonian
$H$ and the corresponding curve $V$.
We can transform $V$ into $V_{0}$, which is a direct sum of
the normal forms, using the gauge transformation in $G$:
\begin{equation}
V_{0}(t) = S(t)V(t)S(0)^{-1}.
\end{equation}
Then we can choose a normal form of $V$ as
\begin{equation}
V_{k} = V_{0}*U_{k}.
\end{equation}
Here, $k$ denotes the winding number of $S$, and
$U_{k}$ is a closed curve which start from the origin and
have the winding number $k$. (See Fig.\ref{normal}.) 
We can choose a normal form
of $U_{k}$ as, for example, 
\begin{equation}
U_{k}(t) = \left(
\begin{array}{ccc}
\left(\begin{array}{cc}
\cos 2\pi kt & - \sin 2\pi kt  \\
\sin 2\pi kt &   \cos 2\pi kt 
\end{array}\right)
& & \\
& \begin{array}{cc}
1 & \\
  & .
\end{array}
& \\
& & \begin{array}{cc}
. & \\
  & 1
\end{array}
\end{array}
\right).
\end{equation}
We use this normal form $V_{0}*U_{k}$ as a representative of $F/\Gzerosim$.
The corresponding Hamiltonian is
\begin{equation}
H(\bp,\bq,t) = \left\{
\begin{array}{cc}
2k\pi(p_{1}^{2}+q_{1}^{2}) & (0\le t < 1/2), \\
H_{0}(\bp,\bq,2t-1) & (1/2\le t \le 1)
\end{array}
\right. 
\end{equation}
Here, $H_{0}$ is the Hamiltonian corresponding to $V_{0}$.

\subsection{Absolute values and Maslov indices of quadratic
path integrals}\label{absolu}

We can reduce a curve $V$ to the normal form $V_{0}*U_{k}$ 
by means of a gauge transformation in $G_{0}$. 
The absolute value and the phase
(Maslov index) of the path integral are the invariants of this 
transformation, and we can easily calculate the path integral
of this normal form. 

The absolute value of the path integral corresponding to
$V_{0}*U_{k}$ is the same as that of $V_{0}$, and the 
Maslov index of $V_{0}*U_{k}$ is $(\mbox{index of }V_{0})+2k$.   
Since $V_{0}$ is a direct sum of the normal forms 
shown in \ref{G0}, the calculation of the path integral 
is reduced to the calculation of the path integrals of the
normal forms. We summarize the results of the calculations
in Table \ref{summerize}. For details of the calculations, see 
appendix \ref{direct}.
\begin{table}
\begin{center}
\begin{tabular}{|c|c|c|c|c|} \hline
         stability & $\lambda$ & diagonal elements 
& $|\Det\Dslash| $ & index\\ \hline
elliptic   &   $e^{\pm i\alpha}$    & 
$-\alpha + 2m\pi$& $4\sin^{2}\alpha/2$ & 1\\ \hline
hyperbolic &   $e^{\pm \beta}$     & 
$\pm\beta,\pm\sqrt{\beta^{2}+(2m\pi)^{2}}$ & 
$4\sinh^{2}\beta/2$ & 0 \\ \hline
inverse hyperbolic & $-e^{\pm \beta}$  & 
$\pm\sqrt{\beta^{2}+(2m+1)^{2}\pi^{2}}$ &
$4\cosh^{2}\beta/2$ & 1 \\ \hline
loxodromic &   $e^{\pm i\alpha \pm \beta}$ &
$\pm\sqrt{\beta^{2}+(\alpha +2m\pi)^{2}}$ &
$4(\cosh\beta - \cos\alpha)^{2}$ & 0 \\ \hline
parabolic & 1 &
0,$-\gamma$, $-\gamma /2 \pm \sqrt{(\gamma /2)^{2} + (2m\pi)^{2}}$ &
$0\;\; (\Det^{'}\Dslash = -\gamma)$ & $(1/2){\rm sgn}\gamma$ \\ \hline
\end{tabular}
\end{center}
\caption{The absolute values of the functional determinants, diagonal
elements of $\Dslash$, and
the Maslov indices for the normal forms.$m$ runs over all integers
in elliptic and loxodromic cases. In hyperbolic and parabolic cases, 
$m\ge 1$, and in inverse hyperbolic case, $m\ge 0$.}
\label{summerize}
\end{table}

The functional determinant corresponding to $V_{0}$ 
is the product of the functional determinant corresponding to
the normal 
forms, and the Maslov index of $V_{0}$ is the sum of the
Maslov index of the normal forms and $2k$.
Therefore, if the monodromy matrix has no degenerate eigenvalue,
we obtain the formula
\begin{equation}
|\Det\Dslash| = |\det (M-I)|,
\end{equation} 
\begin{equation}
\mu = p + q + 2k,
\end{equation}
where p (q) is the number of elliptic (inverse hyperbolic)
pairs of eigenvalues. The partition function become
\begin{eqnarray}
Z & = & \int {\cal D}\bx \exp 
\left[\frac{i}{2}\bx^{T}\Dslash\bx\right],\\
& = & \frac{e^{-i\frac{\pi}{2}\mu}}{\sqrt{|\det(M-I)|}}.
\end{eqnarray}
If the monodromy matrix has parabolic blocks, we must add
$\sum_{i}\frac{1}{2}{\rm sgn}\gamma_{i}$ to the Maslov index.
In this case, $\Dslash$ has zero-modes, and $\Det\Dslash = 0$.
In the space where the zero-modes are removed, the functional
determinant become
\begin{equation}
|\Det^{'}\Dslash| = \sqrt{|\det(M^{'}-I)|}\sqrt{\prod_{i}|\gamma_{i}|}.
\end{equation} 
Here, $\Det^{'}$ denotes the functional determinant in the space without
zero-modes, and $M^{'}$ is the monodromy matrix whose parabolic parts
are removed. Integration with respect to zero-modes will be discussed 
in \S \ref{zero}. 

\subsection{Repetitions of an orbit}\label{repeti}
\begin{figure}
\epsfxsize = 0.8\textwidth
\centerline{\epsffile{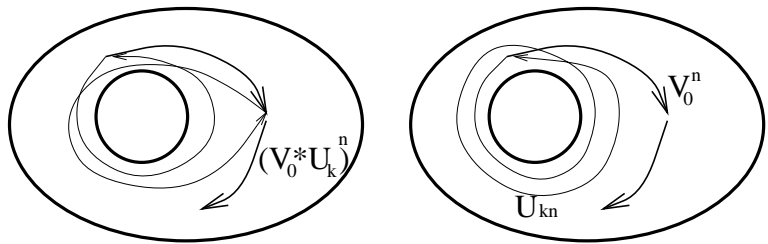}}
\caption{$(V_{0}*U_{k})^{n}$ can be transformed into 
$V_{0}^{n}*U_{2kn}$.}
\label{repetition}
\end{figure}

Let us discuss repetitions of an orbit.
This is the problem of calculating the partition function
\begin{equation}
Z(n) = {\rm Tr}\; \hat{U}(T=n) = {\rm Tr}\{\hat{U}(T=1)^{n}\}.
\end{equation}
If the classical time evolution operator corresponding to 
$Z(1)$ is $V$, that corresponding to $Z(n)$ is 
$V^{n} = V*V*....*V$. Let the normal form of $V$ be $V_{0}*U_{k}$.
Then we can transform $(V_{0}*U_{k})^{n}$ into $V_{0}^{n}*U_{kn}$
continuously. (See Fig. \ref{repetition}.)
Therefore the problem is reduced to the calculation of the repetition
of the normal forms. 

The Maslov index of 
the normal forms of hyperbolic, loxodromic and parabolic type
are unchanged by the repetition.
For elliptic type, the winding number increases when $n\alpha$
goes over $2\pi$, where $n$ is the number of the repetition.
Therefore the Maslov index of this part is $1+[n\alpha/2\pi]$.
The normal form of inverse hyperbolic type can be written as
$V_{h}*U_{\frac{1}{2}}$, where $V_{h}$ is the normal form of
hyperbolic type. Therefore n-repetition of this normal form
can be transformed to $V_{h}^{n}*U_{\frac{n}{2}}$. If $n$ is
even, this is hyperbolic, and if $n$ is odd, this is inverse
hyperbolic type. The Maslov index of this part is $n$.

Thus we obtain the formula for the Maslov index of the 
n-repetition:
\begin{equation}
\mu_{n} = \sum_{i=1}^{p}
\left(1+2\left[\frac{n\alpha_{i}}{2\pi}\right]\right)
+ qn + 2nk.
\label{repet}
\end{equation}
Here, $\alpha_{i}$ denotes the stability angle of i-th
elliptic block. Note that $\mu_{n}\ne n\mu_{1}$ if the orbit
has elliptic blocks.

%% file: g.tex
\section{Trace of the resolvent operator}\label{g}
In this section, we discuss the trace of the resolvent operator
\begin{equation}
g(E) = {\rm Tr} \frac{1}{E-\hat{H}},
\end{equation}
to discuss eigenvalues of the time-independent Hamiltonian
$\hat{H}$.
This is obtained by Fourier transformation of
the partition function:
\begin{equation}
g(E) = \frac{1}{i\hbar} \int_{0}^{\infty} e^{iET/\hbar}Z(T).
\end{equation}
We can find eigenvalues of $\hat{H}$ from the poles of $g(E)$.

$Z(T)$ become a sum over periodic orbits, as discussed in the previous
sections. In this case, each orbit has at least one zero-mode 
corresponding to time-translation symmetry. Therefore we first discuss
the integration with respect to zero-modes in \S \ref{zero}.
Then we carry out Fourier transformation in \S \ref{fourier},
and finally we give a formula for the sum over repetitions of 
a orbit in \S \ref{sum}.

\subsection{Integration with respect to zero-modes}\label{zero}

If the system has continuous symmetries, $\Dslash$ has 
zero-modes and we must execute the integration
with respect to them.  In our formalism based
on the path integral representation of the partition function,
it is easy to treat this integration because the integrand 
$(-i\det^{'}\Dslash)^{-1/2}
\exp \left[\frac{i}{\hbar}R\right]$
is a constant. Therefore 
the contribution from a continuous family of orbits is 
\begin{equation}
\frac{V}{\sqrt{-i\det^{'}\Dslash}}
\exp \left[\frac{i}{\hbar}R\right],
\end{equation}
where $V$ is a volume factor of the family of orbits.

Hereafter, we consider an isolated orbits in a time-independent
Hamiltonian system. In this case, the orbit has a parabolic
block corresponding to the time translation symmetry. 

The parabolic block become
\cite{creagh}\footnote{Note that the notation of the vector in
phase space is different from \cite{creagh}. In this paper, we note
$p$ (or $E$ in this case) first, and $q$ (or $t$) second. This is
opposite of the notation in \cite{creagh}.}
\begin{equation}
m = \left(
\begin{array}{cc}
1 & 0 \\
-\frac{dT}{dE} & 1
\end{array}\right),
\label{parab}
\end{equation} 
which is conjugate to
\begin{equation}
\left(
\begin{array}{cc}
1 & \frac{dT}{dE}\\
0 & 1
\end{array}\right)
\end{equation}
Therefore the contribution from this block is
\begin{equation}
\frac{T}{\sqrt{2\pi\hbar \frac{1}{i}\frac{dT}{dE}}},
\end{equation}
(See appendix \ref{direct}.)
and the amplitude factor of this orbit is
\begin{equation}
K = \frac{T e^{-i\frac{\pi}{2}\mu}}
{\sqrt{2\pi\hbar \frac{dT}{dE}}\sqrt{|\det(M^{'}-I)|}}.
\label{K}
\end{equation}
Here, $M^{'}$ denotes the monodromy matrix whose parabolic part
is removed, and the Maslov index $\mu$ become
\begin{equation}
\mu = p + q + 2k 
- \frac{1}{2}{\rm sgn}\left(\frac{dT}{dE}\right).
\end{equation}

\subsection{Conversion from $T$ to $E$}\label{fourier}

Let us discuss the conversion from a fixed period $T$
to a fixed energy $E$.
The partition function for a fixed $T$ is 
approximated by a sum over periodic orbits as
\begin{equation}
Z(T) = \sum_{p.o.}K\exp 
\left[\frac{i}{\hbar}\oint \bp d\bq - Hdt\right],
\end{equation} 
where the amplitude factor $K$ is given in (\ref{K}).
The trace of the resolvent $g(E)$ is the Fourier transformation
of $Z(T)$:
\begin{equation}
g(E) = \int_{0}^{\infty} dT\; K\exp 
\left[\frac{i}{\hbar}\left\{ET +\oint (\bp d\bq - H dt)\right\}\right].
\end{equation}
Let us evaluate this integral by the stationary phase approximation.
The stationary phase condition
\begin{equation}
\frac{d}{dT} \left\{ET + \oint (\bp d\bq - Hdt)\right\}=0
\end{equation}
leads to 
\begin{equation}
H(\bp(t),\bq(t)) = E.
\end{equation}
The second derivative of the phase is
\begin{eqnarray}
\frac{d^{2}}{dT^{2}} \left\{ET + \oint (\bp d\bq - Hdt)\right\}
& = & - \frac{dE}{dT},
\end{eqnarray}
and the contribution from this factor cancel the contribution
from the parabolic block (\ref{parab}). Therefore $g$ is
approximated as
\begin{equation}
g(E) = \sum_{p.o.}K^{'}\exp\left[\frac{i}{\hbar}S\right],
\end{equation}
\begin{equation}
S = \oint \bp d\bq,
\end{equation}
\begin{equation}
K^{'} = \frac{T e^{-i\frac{\pi}{2}\mu^{'}}}
{\sqrt{|\det (M^{'}-I)|}},
\end{equation}
\begin{equation}
\mu^{'} = p + q + 2k.
\end{equation}
This is the well-known Gutzwiller's trace formula.

\subsection{Sum over repetitions}\label{sum}

\subsubsection{One dimensional case}
\begin{figure}
\epsfxsize = 0.6\textwidth
\centerline{\epsffile{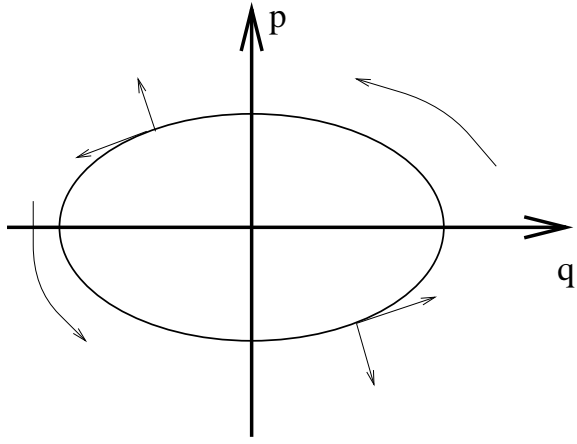}}
\caption{The periodic orbit of the one-dimensional system.
The local coordinate around the orbit rotate once in the phase
space when the orbit goes around the course once.}
\label{1dim}
\end{figure}

Let us discuss the sum over repetitions of a orbit using the
formula (\ref{repet}). First, we consider the one dimensional
case. In this case, there is only one type of periodic orbit.
The monodromy Matrix has only one block 
corresponds to time translation symmetry, and the winding number
of the orbit is 1. (See Fig. \ref{1dim}.) Therefore the Maslov 
index of the prime periodic orbit is $2$ 
and the index of the n-repetition
is $2n$.
The trace of the resolvent $g$ is approximated as
\begin{eqnarray}
g(E) & = & g_{0}(E) + \sum_{n=1}^{\infty} 
T\exp \left[n\left(\frac{i}{\hbar}S(E) - i\pi\right)\right],\\
& = &
g_{0}(E)+T\frac{\exp i[S(E)/\hbar-\pi]}{1-\exp i[S(E)/\hbar-\pi]}.
\end{eqnarray}
Here, $g_{0}(E)$ is the Thomas-Fermi term 
which represents the contribution
of the zero-length orbit.
$T$ and $S$ are the period and the action of the prime 
periodic orbit.

Since the poles of $g(E)$ represent the eigenvalues of the Hamiltonian,
we obtain the Bohr-Sommerfeld quantization condition:
\begin{equation}
S(E) = \oint pdq = (2n+1)\pi\hbar.
\end{equation}

\subsubsection{Elliptic orbits}
 
Let us consider the periodic orbits where all eigenvalues of
the monodromy matrix is elliptic except the parabolic block
corresponds to the time-translation symmetry.
In this case, 
\begin{equation}
|\det (M^{'}-I)| = \prod_{j=1}^{p}
\left|2\sin\frac{\alpha_{j}}{2}\right|,
\end{equation}
where $p$ is the number of elliptic blocks, and $\alpha_{j}$ is
the stability angle of the j-th elliptic block.
The Maslov index is 
\begin{equation}
\mu = p + 2k,
\end{equation}
where $k$ is the winding number. Hence the amplitude factor 
of this orbit is
\begin{equation}
K = \frac{(-1)^{k}T}{\prod_{j=1}^{p}2i\sin\frac{\alpha_{j}}{2}}.
\end{equation}
From the formula (\ref{repet}), we obtain
\begin{equation}
\mu_{n} = \sum_{j=1}^{p}
\left(1+2\left[\frac{n\alpha_{j}}{2\pi}\right]\right)
+ 2kn
\end{equation}
and
\begin{equation}
K_{n} = \frac{(-1)^{kn}T}{\prod_{j=1}^{p}2i\sin\frac{n\alpha_{j}}{2}}
\end{equation}
for n-repetition of this orbit.

Therefore the sum over repetitions of this orbit is
\begin{eqnarray}
\sum_{n=1}^{\infty} K_{n}
\exp\left[\frac{i}{\hbar}nS(E)\right] & = &
T\sum_{n-1}^{\infty}(-1)^{kn}\left\{\prod_{j=1}^{p}\sum_{m_{j}=0}^{\infty}
e^{-i\left(m_{j}+\frac{1}{2}\right)n\alpha_{j}}\right\}
\exp\left[\frac{i}{\hbar}nS(E)\right]\\
& = &
T\sum_{n=1}^{\infty}\sum_{\{m_{j}\}} 
\exp \frac{in}{\hbar}\left[S(E)- \left\{k\pi + 
\sum_{j=1}^{p}\left(m_{j}+\frac{1}{2}\right)\alpha_{j}\right\}\hbar
\right],\\
& = &
T\sum_{\{m_{j}\}}\frac{\exp \frac{i}{\hbar}
\left[S(E)-\left\{k\pi + 
\sum_{j=1}^{p}\left(m_{j}+\frac{1}{2}\right)\alpha_{j}\right\}\hbar
\right]}{1-\exp \frac{i}{\hbar}\left[S(E)-\left\{k\pi + 
\sum_{j=1}^{p}\left(m_{j}+\frac{1}{2}\right)\alpha_{j}\right\}\hbar
\right]}.
\end{eqnarray}
Here, we used
\begin{equation}
\frac{1}{\sin \frac{x}{2}} = 2i\sum_{m=0}^{\infty}
e^{-i\left(m+\frac{1}{2}\right)x}.
\end{equation}
Therefore the quantization condition is
\begin{equation}
S(E) = \left\{(2m_{0}+k)\pi + \sum_{j=1}^{n-1}
\left(m_{j}+\frac{1}{2}\right)\alpha_{j}
\right\}\hbar,
\label{quant}
\end{equation}
where $n$ is an arbitrary integer and
$m_{j}$ is an arbitrary integer which is not negative.
Note that the quantization condition depends on if the winding
number $k$ is even or odd.

\subsubsection{Unstable orbits}

Let the monodromy matrix of a periodic orbit have $p$ elliptic 
blocks, $q$ hyperbolic blocks, $r$ inverse hyperbolic blocks,
$s$ loxodromic blocks and a parabolic block corresponding to the
time translation symmetry. Then the eigenvalues of the monodromy
matrix can be written as $\exp (\pm i\zeta )$ using the stability
angle $\zeta$, and $\zeta$ is represented as 
\begin{eqnarray}
\zeta_{e,j_{e}} & = & \alpha_{e,j_{e}}\;\;\; (1\le j_{e}\le p),
\label{el}\\
\zeta_{h,j_{h}} & = & i\beta_{h,j_{h}}\;\;\; (1\le j_{h}\le q),\\
\zeta_{ih,j_{ih}} & = & i\beta_{ih,j_{ih}}+\pi \;\;\; (1\le j_{ih}\le r),\\
\zeta_{l,j_{l},\pm} & = & \pm\alpha_{l,j_{l}} + i\beta_{l,j_{l}},
(1\le j_{l}\le s) \label{lox}\\
\zeta_{p} & = & 1.
\end{eqnarray}
Here, $0 < \alpha_{e,j_{e}} < 2\pi$ and $0<\alpha_{l,j_{l}}<\pi$.
$\beta_{h,j_{h}}$, $\beta_{ih,j_{ih}}$ and $\beta_{l,j_{l}}$
are positive. $p + q + r + 2s + 1 = n$ where $2n$ is the dimension of
the phase space. 

The amplitude factors of this orbit and its repetitions are
\begin{equation}
K_{n} = (-1)^{kn}T K_{e,n}K_{h,n}K_{ih,n}K_{l,n},
\end{equation}
where
\begin{eqnarray}
K_{e,n} & = & \frac{1}{\prod_{j=1}^{p}2i\sin\frac{n\alpha_{e,j}}{2}},\\
K_{h,n} & = & \frac{1}{\prod_{j=1}^{q}2\sinh\frac{n\beta_{h,j}}{2}},\\
K_{ih,n} & = & \left\{
\begin{array}{cc}
\frac{(-1)^{(n-1)/2}}{\prod_{j=1}^{r}2i\cosh\frac{n\beta_{ih,j}}{2}} & 
(\mbox{$n$: odd})\\
\frac{(-1)^{n/2}}{\prod_{j=1}^{r}2\sinh\frac{n\beta_{ih,j}}{2}} &
(\mbox{$n$: even})
\end{array}\right.\\
K_{l,n} & = & \frac{1}{\prod_{j=1}^{s}
2(\cosh n\beta_{l,j_{l}}-\cos n\alpha_{l,j_{l}})}.
\end{eqnarray}
The sum over repetitions of this orbit can be calculated in the same
way as elliptic orbits, and the result is
\begin{equation}
\sum_{n=1}^{\infty}K_{n}\exp\left[\frac{i}{\hbar}nS(E)\right] =
T\sum_{\bm}
\frac{\exp\left[\frac{i}{\hbar}S_{c}(E,\bm)\right]}
{1-\exp\left[\frac{i}{\hbar}S_{c}(E,\bm)\right]},
\end{equation}
where
\begin{equation}
S_{c}(E,\bm ) = S(E) - \left\{k\pi + \sum_{j=1}^{n-1}
\left(m_{j}+\frac{1}{2}\right)\zeta_{j}\right\}\hbar
\end{equation}
Here, $\bm = (m_{1},m_{2},...,m_{n-1})$ is the $n-1 (= p+q+r+2s)$ 
dimensional vector whose components are non-negative integers. 
$\zeta_{j}$ represents one of stability angles in
Eqs. (\ref{el})-(\ref{lox}). Therefore the quantization condition
is
\begin{equation}
S(E) = \left\{(2m_{0}+k)\pi + \sum_{j=1}^{n-1}
\left(m_{j}+\frac{1}{2}\right)\zeta_{j}\right\}\hbar,
\label{quant2}
\end{equation}
where $m_{0}$ is an arbitrary integer. If some of stability angles
have imaginary parts, RHS of (\ref{quant2}) have imaginary parts
and eigenvalues become complex numbers.

In chaotic systems, most of the periodic orbits are unstable
and the 
eigenvalues determined by the condition (\ref{quant2})
also have imaginary parts.
Such phenomena occur even if the Hamiltonian of the system is 
a Hermite operator and leads to false results.

This means that, for unstable orbits, it is not enough to consider
only one orbit and its repetitions. We should take into account 
the correlations of unstable orbits. This is a difficult but
interesting future problem.

%% file: concl.tex
\section{Conclusion}

We have investigated the semiclassical trace formula using phase space 
path integral. We derived the semiclassical trace formula by 
applying the stationary approximation to the phase space path
integral of the partition function.

We classified the quadratic path integrals around the periodic
orbit. This is equivalent to the classification of connections on 
the vector bundle over $S^{1}$ with the structure group $Sp(2n,R)$.
However, if the canonical transformation is topologically
non-trivial, the Maslov index changes.
Therefore we should regard two connections as equivalent
only if they are connected by the topologically trivial
canonical transformation. The results of the classification
by this equivalence relation is shown in \S \ref{G0}.

We defined normal forms of time-dependent quadratic Hamiltonians,
and calculated quadratic path integrals. We also derived a
formula for the Maslov index of the n-repetition of the orbit 
in \S \ref{repeti},
and derived the Bohr-Sommerfeld type quantization condition
in \S \ref{sum}.

Throughout this paper, we concentrated on the quadratic path integrals
around periodic orbits. However, the analysis of higher order terms
is often needed when we investigate realistic systems like
nuclei and atomic clusters. Phase space path integrals 
used in this paper will
supply powerful tools for such analysis.

\section*{Acknowledgment}
The author would like to thank Professor T.Hatsuda for helpful
discussions and encouragements. He would also like to thank   
Professor M.Sano,
Dr. S.Sugimoto, 
Professor T.Kunihiro, Professor H.Kuratsuji, Professor A.G.Magner
and the members of
the Nuclear Theory Group at Kyoto University
for valuable discussions.

%% file: diag.tex
\section{Calculations of diagonal elements in the continuum
limit}\label{diag}
\subsection{Diagonalization of $D$}\label{diagD}
First we consider the following eigenvalue equation:
\begin{equation}
D\bx = \left(\frac{d}{dt}+A(t)\right) \bx = \epsilon \bx.
\end{equation}
This equation is easy to solve because the time derivative has 
diagonal form.
Let $\bxi_{\zeta}$ be the eigenvector of the monodromy matrix $M$,
and $\zeta$ be the stability angle:
\begin{equation}
M\bxi_{\zeta} = e^{-i\zeta}\bxi_{\zeta}.
\label{eigenM}
\end{equation}
Then the solutions of (\ref{eigenM}) are written as follows:
\begin{equation}
\bx_{\zeta,m}(t) = \exp(\epsilon_{\zeta,m}t)V(t)\bxi_{\zeta},
\end{equation}
\begin{equation}
D\bx_{\zeta,m} = \epsilon_{\zeta,m}\bx_{\zeta,m},
\end{equation}
\begin{equation}
\epsilon_{\zeta,m} = (\zeta + 2m\pi)i.
\label{eigD}
\end{equation}
Here, $V(t)$ is the classical time evolution operator
which satisfies
\begin{equation}
DV(t) = 0.
\end{equation}

We can calculate $|\Det\Dslash|=|\Det D|$ from (\ref{eigD}):
\begin{eqnarray}
|\Det D| & = & \left|\prod_{l=1}^{n}\prod_{m=-\infty}^{\infty}
(2m\pi +\zeta_{l})(2m\pi -\zeta_{l})\right|,\\
& = & \left|\prod_{l=1}^{n}\prod_{m=-\infty}^{\infty}
(2m\pi)^{2}-\zeta_{l}^{2}\right|,\\
& = & \left|\prod_{l=1}^{n}\zeta_{l}^{2}\prod_{m=1}^{\infty} (2m\pi)^{2}
\left\{1-\left(\frac{\zeta_{l}}{2m\pi}\right)^{2}\right\}^{2}
\right|,\\
& \sim & 
\left|\prod_{l=1}^{n} \left(2\sin\frac{\zeta_{l}}{2}\right)^{2}\right|
= \left|\det (M-I)\right|.
\end{eqnarray}
Here, we used the formula (\ref{sin}) and ignored the diverging constant
$\prod_{m=1}^{\infty}2m\pi$. We can see from the calculations in Appendix \ref{direct} that this constant corresponds
to 
\begin{equation}
\frac{1}{N}\prod_{m=1}^{N-1}2\sin\frac{\pi m}{N} = 1
\end{equation}
in the discrete formalism. 

\subsection{Diagonalization of $\Dslash$}
In this subsection, we diagonalize $\Dslash$ using the
basis given in \ref{diagD}.

\subsubsection{Elliptic type}\label{elliptic}
Eigenvalues of the monodromy matrix are
\begin{equation}
M\bx_{\pm} = e^{\mp i\alpha}\bx_{\pm} \;\;\;(0 < \alpha < 2\pi),
\end{equation}
$(\bx_{+} = \bx^{*}_{-})$. 
$\alpha$ is taken $0 < \alpha < \pi$
if $[\bx,M\bx]>0$, and $\pi <\alpha < 2\pi$ if $[\bx,M\bx]<0$, 
\footnote{The product $[\bx ,\by]$ means $\bx^{T}J \by$.}
where
$\bx $ is an arbitrary real vector in the space spanned by $\bx_{+}$
and $\bx_{-}$.  See also \cite{arnold} (\S 42).
We rewrite $\bx_{+}$ as $\bx_{+}=\bx_{Re}+ i\bx_{Im}$ where $\bx_{Re}$ and
$\bx_{Im}$ are real vectors, then we obtain
\begin{eqnarray}
M \bx_{Re} & = & \bx_{Re}\cos\alpha + \bx_{Im}\sin\alpha ,\\
M \bx_{Im} & = & \bx_{Re}\sin\alpha - \bx_{Im}\cos\alpha .
\end{eqnarray}
The inequality 
\begin{equation}
[\bx_{Re},\bx_{Im}] > 0
\end{equation}
follows from the definition of $\alpha$.
We use the normalization condition 
\begin{equation}
[\bx_{Re},\bx_{Im}] = 1, 
\end{equation}
\begin{equation}
[\bx_{+},\bx_{-}] = - 2i .
\end{equation}

The eigenfunctions of $D$ are
\begin{equation}
\bx_{\pm,m}(t) = \exp (\epsilon_{\pm,m}t)V(t)\bx_{\pm},
\end{equation}
\begin{equation}
\epsilon_{\pm,m} = (\pm\alpha + 2m\pi)i.
\end{equation}
We define the real functions $\bx_{Re,m},\bx_{Im,m}$ as
\begin{eqnarray}
\bx_{Re,m} & = & \frac{1}{2}(\bx_{+,m}+\bx_{-,-m}), \\
\bx_{Im,m} & = & \frac{1}{2i}(\bx_{+,m}-\bx_{-,-m})\;\;\;(n=0,\pm
1,\pm2,....) .
\end{eqnarray}
These functions satisfy the normalization and the orthogonalization
condition:
\begin{eqnarray}
\int_{0}^{1}\! dt \: [\bx_{Re,m},\bx_{Im,n}]  =  \delta_{m,n}, \\
\int_{0}^{1}\! dt \: [\bx_{Re,m},\bx_{Re,n}]  = 
\int_{0}^{1}\! dt \: [\bx_{Im,m},\bx_{im,n}]  =  0 .
\end{eqnarray}
Matrix elements of $\Dslash$ are
\begin{eqnarray}
\int_{0}^{1}dt \bx_{Re,m}^{T} \Dslash \bx_{Re,n} & = & 
- (\alpha + 2m\pi)\delta_{m,n}, \\
\int_{0}^{1}dt \bx_{Im,m}^{T} \Dslash \bx_{Im,n} & = & 
- (\alpha + 2m\pi)\delta_{m,n}, \\
\int_{0}^{1}dt \bx_{Re,m}^{T} \Dslash \bx_{Im,n} & = & 0 .
\end{eqnarray}
Therefore $\Dslash$ is already diagonalized by this basis.

The functional determinant of this part is 
\begin{eqnarray}
\prod_{m=-\infty}^{\infty} (\alpha + 2m\pi)^{2}
& = & \left\{\prod_{m=1}^{\infty}(2m\pi)^{4}\right\}
      \alpha^{2}\prod_{m=1}^{\infty}
      \left\{1 - \left(\frac{\alpha}{2m\pi}\right)^{2}\right\}, \\
& \sim & 4\sin^{2}\frac{\alpha}{2} .
\end{eqnarray}
Here we used the formula (\ref{sin}).

\subsubsection{Hyperbolic type}\label{hyperbolic}
In this case, the eigenvectors of the monodromy matrix satisfy
the equation
\begin{equation}
M \bx_{\pm} = e^{\pm \beta}\bx_{\pm}
\end{equation}
where $\beta$ is a real number.
The normalization condition is taken as
\begin{equation}
[\bx_{+},\bx_{-}]=1.
\end{equation}

We define $\epsilon_{\pm,m},\bx_{\pm,m}(t)$ as before:
\begin{equation}
\bx_{\pm,m} = \exp (\epsilon_{\pm,m}t)V(t)\bx_{\pm},
\end{equation}
\begin{equation}
\epsilon_{\pm,m} = \mp \beta + 2m\pi i.
\end{equation}
We define real functions
$\bx_{c,\pm,m},\bx_{s,\pm,m}$ as follows:
\begin{eqnarray}
\bx_{c,\pm,m} & = & \frac{1}{\sqrt{2}} (\bx_{\pm,m} + \bx_{\pm,-m}), \\
\bx_{s,\pm,m} & = & \frac{1}{i\sqrt{2}}(\bx_{\pm,m} - \bx_{\pm,-m}) \;\;\; 
(m=1,2,...) .
\end{eqnarray}
Since $\bx_{\pm,0}$ are already real, 
we don't have to redefine them.
Matrix elements which have non-zero values are
\begin{eqnarray}
\int_{0}^{1}\! dt \; \bx_{c,+,m}^{T} \Dslash \bx_{c,-,m} 
& = & \beta ,\\
\int_{0}^{1}\! dt \; \bx_{s,+,m}^{T} \Dslash \bx_{c,-,m}
& = & - 2m\pi ,\\
\int_{0}^{1}\! dt \; \bx_{c,+,m}^{T} \Dslash \bx_{s,-,m}
& = & 2m\pi ,\\
\int_{0}^{1}\! dt \; \bx_{s,+,m}^{T} \Dslash \bx_{c,-,m}
& = & \beta ,\\
\int_{0}^{1}\! dt \; \bx_{+,0}^{T} \Dslash \bx_{-,0}
& = & \beta .
\end{eqnarray}
Therefore $\Dslash$ is represented by the matrix $d_{m}$ 
in the space spanned by 
$\bx_{c,\pm,m},\bx_{s,\pm,m}$: 
\begin{equation}
d_{m} = 
\left(
\begin{array}{cccc}
0 & 0 & \beta & 2m\pi\\
0 & 0 & -2m\pi &   \beta\\
\beta &  -2m\pi & 0 & 0 \\
2m\pi & \beta & 0 & 0
\end{array}
\right)\;\;\;(m\ge 1).
\end{equation}
In the case $m=0$,
\begin{equation}
d_{0} =
\left(
\begin{array}{cc}
0 & \beta \\
\beta & 0 \\
\end{array}
\right) .
\end{equation}
The solutions of the eigenvalue equation $\det (d_{m}-\lambda I)=0$
are
\begin{eqnarray}
\lambda & = & \pm \sqrt{\beta^{2} + (2m\pi)^{2}} \;\;\;\;(m\ge 1), \\
\lambda & = & \pm \beta \;\;\;\; (m=0) .
\end{eqnarray}
The solutions for $m\ge 1$ are doubly degenerate.

The functional determinant of this part is (using (\ref{sinh}))
\begin{eqnarray}
\prod_{m=0}^{\infty}\det d_{m}
& = & \beta^{2}\prod_{m=1}^{\infty}
\left\{\beta^{2} + (2m\pi)^{2}\right\}^{2}, \\
& \sim & 4\sinh^{2}\frac{\beta}{2} .
\end{eqnarray}

\subsubsection{Inverse hyperbolic type}
In this case, eigenvalues of the monodromy matrix are negative
real numbers:
\begin{equation}
M \bx_{\pm} = - e^{\pm \beta}\bx_{\pm}.
\end{equation}
The normalization condition is taken as
\begin{equation}
[\bx_{+},\bx_{-}] = 1.
\end{equation}

Eigenvalues of $D$ are
\begin{equation}
\epsilon_{\pm,m} = \mp \beta + (2m+1)\pi i.
\end{equation}
We define real vectors $\bx_{c,\pm,m},\bx_{s,\pm,m}$ as follows:
\begin{eqnarray}
\bx_{c,\pm,m} & = & \frac{1}{\sqrt{2}} (\bx_{\pm,m} + \bx_{\pm,-m-1}) \\
\bx_{s,\pm,m} & = & \frac{1}{i\sqrt{2}}(\bx_{\pm,m} - \bx_{\pm,-m-1})
\;\;\;(m=0,1,2,....)  
\end{eqnarray}
The matrix elements of $\Dslash $ can be calculated in the same way as
the hyperbolic case.
In the space spanned by $\bx_{c,\pm,m}$ and $\bx_{s,\pm,m}$,
$\Dslash$ is represented as
\begin{equation}
d_{m} =
\left(
\begin{array}{cccc}
0 & 0 & \beta & (2m+1)\pi\\
0 & 0 & - (2m+1)\pi & \beta\\
\beta &  - (2m+1)\pi & 0 & 0 \\
(2m+1)\pi & \beta & 0 & 0
\end{array}
\right)
\end{equation}
Eigenvalues of $d_{m}$ are
\begin{equation}
\lambda = \pm \sqrt{\beta^{2}+(2m+1)^{2}\pi^{2}}.
\end{equation}
Each solution is doubly degenerate.
The functional determinant of this part is 
\begin{eqnarray}
\prod_{m=0}^{\infty}d_{m} & = &
\prod_{m=0}^{\infty}\left\{\beta^{2}+(2m+1)^{2}\pi^{2}\right\}^{2}\\
& \sim & 4\cosh^{2}\frac{\beta}{2}.
\end{eqnarray}
Here we used (\ref{cosh}).

\subsubsection{Loxodromic type}\label{loxodromic}
Eigenvalues of the monodromy matrix are
\begin{equation}
M \bx_{\pm,\pm} = e^{\mp i\alpha \pm \beta}\bx_{\pm,\pm} 
\;\;\;\;(\bx_{\pm,\pm}=\bx_{\mp,\pm}^{*}).
\end{equation}

Normalization is taken as
\begin{equation}
[\bx_{+,+},\bx_{-,-}] = - 2i 
\end{equation}
\begin{equation}
[\bx_{+,-},\bx_{-,+}] = - 2i
\end{equation}
The eigenvectors of $D$ are
\begin{equation}
\bx_{\pm,\pm,m}(t) = \exp (\epsilon_{\pm,\pm,m}t)V(t)\bx_{\pm,\pm},
\end{equation}
where
\begin{equation}
\epsilon_{\pm,\pm,m} = \pm i\alpha \mp \beta + 2m\pi i. 
\end{equation}
We define real basis $\bx_{Re,\pm,m},\bx_{Im,\pm,m}$ as
\begin{eqnarray}
\bx_{Re,\pm,m} & = & \frac{1}{2} (\bx_{+,\pm,m} + \bx_{-,\pm,-m}), \\
\bx_{Im,\pm,m} & = & \frac{1}{2i}(\bx_{+,\pm,m} - \bx_{-,\pm,-m}) 
\;\;\;\;(m=0,\pm1,\pm 2,...).
\end{eqnarray}
These vectors are normalized as
\begin{equation}
[\bx_{Re,+,m},\bx_{Im,-,n}] = 
[\bx_{Re,-,m},\bx_{Im,+,n}] = \delta_{m,n}.
\end{equation}
Other combinations are equal to zero.

The matrix elements which have non-zero values are
\begin{eqnarray}
\int_{0}^{1}\! dt \; \bx_{Re,+,m}^{T} \Dslash \bx_{Re,-,m} 
& = &  - (\alpha + 2m\pi),\\
\int_{0}^{1}\! dt \; \bx_{Re,+,m}^{T} \Dslash \bx_{Im,-,m} 
& = &  \beta ,\\
\int_{0}^{1}\! dt \; \bx_{Im,+,m}^{T} \Dslash \bx_{Re,-,m} 
& = &  -\beta ,\\
\int_{0}^{1}\! dt \; \bx_{Im,+,m}^{T} \Dslash \bx_{Im,-,m} 
& = &  - (\alpha + 2m\pi).
\end{eqnarray}
Therefore $\Dslash$ is represented in the space spanned by
$x_{Re,\pm,m},x_{Im,\pm,m}$ as
\begin{equation}
d_{m} =
\left(
\begin{array}{cccc}
0 & 0 & - (\alpha + 2m\pi) & \beta \\
0 & 0 & -\beta & - (\alpha + 2m\pi) \\
- (\alpha + 2m\pi) & -\beta & 0 & 0 \\
\beta & - (\alpha + 2m\pi) & 0 & 0 
\end{array}
\right).
\end{equation}
Eigenvalues of $d_{m}$ are
\begin{equation}
\lambda = \pm \sqrt{\beta^{2} + (\alpha + 2m\pi)^{2}}.
\end{equation}
Since the determinant of $d_{m}$ is 
$\left\{\beta^{2} + (\alpha + 2m\pi)^{2}\right\}^{2}$, the
functional determinant of this part is
\begin{eqnarray}
\prod_{m=-\infty}^{\infty} \left\{\beta^{2} + (\alpha + 2m\pi)^{2}\right\}^{2}
& = & 
\prod_{m=-\infty}^{\infty}(\alpha + 2m\pi)^{4}
\prod_{m=-\infty}^{\infty}
\left\{1 + \frac{\beta^{2}}{(\alpha +2m\pi)^{2}}\right\}^{2}\\
& \sim & 4(\cosh \beta - \cos \alpha)^{2}.
\end{eqnarray}
Here, we used the formula (\ref{sin}) and (\ref{chcos}).

\subsubsection{Parabolic type}\label{zeromode}
If there are eigenvalues of the monodromy matrix which are equal to 1,
the operator $\Dslash$ has zero-modes.

Here, we treat the most important case, the parabolic type. 
In this case, there are bases $\bx_{1},\bx_{2}$ which satisfy
\begin{eqnarray}
M \bx_{1} & = & \bx_{1}, \\
M \bx_{2} & = & \bx_{2} - \gamma \bx_{1}.
\end{eqnarray}
Here, the normalization is taken as
\begin{equation}
[\bx_{1},\bx_{2}] = 1.
\end{equation}
The monodromy matrix is represented in this space as
\begin{equation}
\left(
\begin{array}{cc}
1 & -\gamma \\
0 & 1
\end{array}
\right).
\end{equation}

We define a matrix $B$ which satisfies
\begin{eqnarray}
B \bx_{1} & = & 0, \\
B \bx_{2} & = &  -\gamma \bx_{1}.
\end{eqnarray}
This is a generator of $M$. ($e^{B}=M$.) 

We can diagonalize $\Dslash$ using these bases.
$\bx_{1,m},\bx_{2,m}$ are defined as
\begin{eqnarray}
\bx_{\alpha,m}(t) & = & 
\exp (2m\pi it)V(t)\exp (-Bt)\bx_{\alpha} 
\;\;\;(\alpha = 1,2).
\end{eqnarray}
$D$ acts on these bases as
\begin{eqnarray}
D \bx_{1,m} & = & 2m\pi i  \bx_{1,m}, \\
D \bx_{2,m} & = & 2m\pi i  \bx_{2,m} + \gamma \bx_{1,m}.
\end{eqnarray}
We define real bases $\bx_{c,\alpha,m},\bx_{s,\alpha,m}$ as
\begin{eqnarray}
\bx_{c,\alpha,m} & = & \frac{1}{\sqrt{2}} (\bx_{\alpha,m} + \bx_{\alpha,-m})\\
\bx_{s,\alpha,m} & = & \frac{1}{i\sqrt{2}}(\bx_{\alpha,m} - \bx_{\alpha,-m})
\;\;\;\;(m=1,2,...).
\end{eqnarray}
$\bx_{1,0}$ and $\bx_{2,0}$ are already real.

Then the matrix element which is not equal zero are
\begin{eqnarray}
\int_{0}^{1}\! dt \: \bx_{c,1,m}^{T} \Dslash \bx_{s,2,m}
& = & 2m\pi, \\
\int_{0}^{1}\! dt \: \bx_{s,1,m}^{T} \Dslash \bx_{c,2,m}
& = & - 2m\pi, \\
\int_{0}^{1}\! dt \: \bx_{c,2,m}^{T} \Dslash \bx_{c,2,m}
& = & - \gamma, \\
\int_{0}^{1}\! dt \: \bx_{s,2,m}^{T} \Dslash \bx_{s,2,m}
& = & - \gamma.
\end{eqnarray}
Therefore, for $m\ge 1$,
\begin{equation}
d_{m} =
\left(
\begin{array}{cccc}
0 & 0 & 0 & 2m\pi \\
0 & 0 & -2m\pi & 0 \\
0 & -2m\pi & -\gamma & 0 \\
2m\pi & 0 & 0 & -\gamma \\
\end{array}
\right),
\end{equation}
\begin{equation}
\det d_{m} = (2m\pi)^{4}.
\end{equation}
Eigenvalues are
\begin{equation}
\lambda = - \frac{\gamma}{2} 
\pm \sqrt{\left(\frac{\gamma}{2}\right)^{2} + (2m\pi)^{2}}.
\end{equation}
Each solution is doubly degenerate.

For $m=0$,
\begin{equation}
d_{0} =
\left(
\begin{array}{cc}
0 & 0 \\
0 & - \gamma
\end{array}
\right)
\end{equation}
The eigenvalues of this part is
\begin{equation}
\lambda = -\gamma,0.
\end{equation}
Since $\Dslash$ has a zero-mode,
\begin{equation}
\Det \Dslash = 0.
\end{equation}
The functional determinant in the space where the zero-mode
is removed is 
\begin{equation}
{\rm det} 'D = - \gamma.
\end{equation}
The zero-mode should be integrated separately.

%% file: direct.tex
\section{Direct calculations of some path integrals}\label{direct}
\subsection{Elliptic type}

Let us consider the normal form of the Hamiltonian of elliptic
type:
\begin{equation}
H = \frac{\alpha}{2}(p^{2}+q^{2}).
\end{equation}
The path integral for this Hamiltonian can be calculated
directly.
The eigenvalue equation of $\Dslash_{N}$ is
\begin{equation}
- \frac{\alpha}{N} p_{j} + q_{j} - q_{j-1} 
 = \epsilon p_{j} ,
\end{equation}
\begin{equation}
p_{j}-p_{j+1} - \frac{\alpha}{4N}(q_{j-1}+2 q_{j}+q_{j+1})
 =  \epsilon q_{j}.
\end{equation}
Since the Hamiltonian is time-independent, we assume that the
solutions can be written as
\begin{equation}
\left(
\begin{array}{c}
p_{j}\\
q_{j}
\end{array}
\right) = 
\left(
\begin{array}{c}
p \\
q
\end{array}
\right) 
\exp \left(i\frac{2\pi r}{N}j\right)
\;\;\;(r=0,1,...,N-1).
\end{equation}
Then the eigenvalue equation becomes
\begin{equation}
\left(
\begin{array}{cc}
- \frac{\alpha}{N} & 1 - e^{-i\frac{2\pi r}{N}} \\
1 - e^{i\frac{2\pi r}{N}} & 
-\frac{\alpha}{4N}(2 + e^{i\frac{2\pi r}{N}} 
+ e^{-i\frac{2\pi r}{N}})
\end{array}
\right) 
\left(
\begin{array}{c}
p\\
q
\end{array}
\right) =
\epsilon
\left( 
\begin{array}{c}
p\\
q
\end{array}
\right).
\end{equation}
$\epsilon$ satisfies the following quadratic equation:
\begin{equation}
\epsilon^{2} + 
\frac{\alpha}{N} \left(1+\cos^{2}\frac{\pi r}{N}\right)\epsilon +
\left(\frac{\alpha}{N}\right)^{2}\cos^{2}\frac{\pi r}{N}
- 4\sin^{2}\frac{\pi r}{N}
= 0.
\label{quad-eq}
\end{equation}
We can find all eigenvalues by solving this equation.

Let us calculate $\det \Dslash_{N}$.
\begin{eqnarray}
\det \Dslash_{N} 
& = & \prod_{r=0}^{N-1}\left\{
\left(\frac{\alpha}{N}\right)^{2}\cos^{2}\frac{\pi r}{N}
- 4\sin^{2}\frac{\pi r}{N}\right\}, \\
& = & \left(\frac{\alpha}{N}\right)^{2}
\prod_{r=1}^{N-1}\left(- 4\sin^{2}\frac{\pi r}{N}\right)
\prod_{r=1}^{N-1}
\left(1-\frac{\alpha^{2}}{4N^{2}}\cot^{2}\frac{\pi r}{N}\right).
\end{eqnarray}
The first part of the product can be calculated using the formula
(\ref{d_norm}):
\begin{equation}
\prod_{r=1}^{N-1} \sin \frac{\pi r}{N}
= \frac{N}{2^{N-1}}.
\end{equation}
For the second part, if $N$ is large enough,
\begin{eqnarray}
\prod_{r=1}^{N-1}
\left(1-\frac{\alpha^{2}}{4N^{2}}\cot^{2}\frac{\pi r}{N}\right)
& = &
\prod_{r=1}^{\left[\frac{N}{2}\right]}
\left(1-\frac{\alpha^{2}}{4N^{2}}\cot^{2}\frac{\pi r}{N}\right)^{2},\\
& \stackrel{N\rightarrow\infty}{\rightarrow} &
\prod_{r=1}^{\infty}\left\{
1- \left(\frac{\alpha}{2\pi r}\right)^{2}
\right\}^{2}.
\end{eqnarray}
We can calculate this infinite product using the formula
(\ref{sin}):
\begin{equation}
\prod_{n=1}^{\infty}\left(1-\frac{x^{2}}{n^{2}}\right)
= \frac{\sin \pi x}{\pi x}.
\end{equation}
Therefore
\begin{equation}
\lim_{N\rightarrow\infty}|\det \Dslash_{N}|
= 4\sin^{2}\frac{\alpha}{2}.
\end{equation}

We can calculate the Maslov index using the quadratic equation
(\ref{quad-eq}). 
If
\begin{equation}
\left(\frac{\alpha}{N}\right)^{2}\cos^{2}\frac{\pi r}{N}
- 4\sin^{2}\frac{\pi r}{N} < 0,
\end{equation}
the solutions of the quadratic equation have opposite sign.
Therefore this part doesn't contribute to the Maslov index.
The problem is the case where
\begin{equation}
\left(\frac{\alpha}{N}\right)^{2}
\cos^{2}\frac{\pi r}{N}
- 4\sin^{2}\frac{\pi r}{N} > 0.
\label{ineq}
\end{equation}
In this case, the solutions of the quadratic equation
have the same sign. 
If $\alpha$ is positive (negative), the coefficient of the linear term
in (\ref{quad-eq}) is also positive (negative) and both solutions
are negative (positive). Therefore such
pair contribute $+1$ ($-1$) to the Maslov index.

(\ref{ineq}) can be rewritten as
\begin{equation}
\left(\frac{\alpha}{N}\right)^{2}
\left\{1 - 4\left(\frac{N}{\alpha}\right)^{2}\tan^{2}\frac{\pi r}{N}
\right\} > 0 \;\;\; (0\le r \le N-1).
\label{ineq}
\end{equation}
$r=0$ always satisfies this inequality, and if $r\ne 0$ satisfies 
(\ref{ineq}), $N-r$ also satisfies (\ref{ineq}).
If $N$ is large enough and $r < N/2$, (\ref{ineq}) 
is reduced to
\begin{equation}
1 - \left(\frac{2\pi r}{\alpha}\right)^{2} > 0.
\end{equation} 
The number of $r (\ne 0)$ satisfying this inequality is 
$\left[\frac{\alpha}{2\pi}\right]$.
Therefore, the Maslov index is
\begin{equation}
\mu = 1 + 2\left[\frac{\alpha}{2\pi}\right].
\end{equation}

\subsection{Hyperbolic type}

The normal form of the Hamiltonian of the hyperbolic type is
\begin{equation}
H = -\beta pq.
\end{equation}
This case can be treated in the similar way to the elliptic type.
The eigenvalue equation is
\begin{equation}
\left(
\begin{array}{cc}
0 & 1-e^{-i\frac{2\pi r}{N}}+\frac{\beta}{2N}(1+e^{-i\frac{2\pi r}{N}}) \\
1-e^{i\frac{2\pi r}{N}}+\frac{\beta}{2N}(1+e^{i\frac{2\pi r}{N}}) & 0
\end{array}
\right) 
\left(
\begin{array}{c}
p\\
q
\end{array}
\right) =
\epsilon
\left( 
\begin{array}{c}
p\\
q
\end{array}
\right),
\end{equation}
\begin{equation}
\epsilon^{2} - 
\left(\frac{\beta^{2}}{N^{2}}\cos^{2}\frac{\pi r}{N} +
4\sin^{2}\frac{\pi r}{N}\right)
= 0.
\label{heq}
\end{equation}
Therefore
\begin{eqnarray}
|\det \Dslash_{N}| & = &  \prod_{r=0}^{N-1}\left\{
\left(\frac{\beta}{N}\right)^{2}\cos^{2}\frac{\pi r}{N}
+ 4\sin^{2}\frac{\pi r}{N}\right\},\\
& = & \left(\frac{\beta}{N}\right)^{2}
\prod_{r=1}^{N-1}\left(4\sin^{2}\frac{\pi r}{N}\right)
\prod_{r=1}^{N-1}\left(1 + \frac{\beta^{2}}{4N^{2}}
\cot^{2}\frac{\pi r}{N}\right).
\end{eqnarray}
\begin{eqnarray}
\lim_{N\rightarrow\infty}|\det \Dslash_{N}| & = &
\beta^{2}
\prod_{r=1}^{\infty}\left\{1 + \left(\frac{\beta}{2\pi r}\right)^{2}
\right\}^{2},\\
& = & 4\sinh^{2}\frac{\beta}{2}.
\end{eqnarray}
Since the two solutions of (\ref{heq}) have always opposite signs,
the Maslov index of this part is 0.

\subsection{Inverse hyperbolic type}

In this case, the normal form of $V$ is
\begin{equation}
V(t) = \left\{
\begin{array}{cc}
\left(
\begin{array}{cc}
\cos 2\pi t & - \sin 2\pi t \\
\sin 2\pi t & \cos 2\pi t
\end{array} \right)
& (0\le t < 1/2), \\
\left(
\begin{array}{cc}
- e^{\beta(2t-1)} & 0 \\
0 & - e^{-\beta(2t-1)}
\end{array}
\right) & (1/2\le t \le 1)
\end{array}\right.
\end{equation}
We calculate the trace of the quantum time evolution operator
corresponding to this $V$:
\begin{eqnarray}
Z & = & {\rm Tr} M\left(V(t=1)\right),\\
& = &
{\rm Tr}\left\{M\left(V(1,1/2)\right) M\left(V(1/2,0)\right)\right\}.
\end{eqnarray}
Here, $V(t_{1},t_{2})$ denotes the classical propagator from 
$t=t_{1}$ to $t=t_{2}$. Let us evaluate this trace using the
coherent states $|z\rangle = |q+ip\rangle$.
\begin{eqnarray}
Z & = & \int \frac{dz_{1}}{\pi}\frac{dz_{2}}{\pi}
\langle z_{1}| M\left(V(1,1/2)\right) |z_{2}\rangle
\langle z_{2}| M\left(V(1/2,1)\right) |z_{1}\rangle .
\end{eqnarray}
Since $\langle z_{2}| M\left(V(1/2,1)\right) |z_{1}\rangle 
= \langle -z_{2}|z_{1}\rangle e^{-i\pi/2}$,
\begin{equation}
Z = \int \frac{dz}{\pi} \langle z|M\left(V(1,1/2)\right)|-z\rangle.
\end{equation}
Therefore to calculate the path integral of the inverse hyperbolic
type is equivalent to calculate the path integral of the hyperbolic
type with anti-periodic boundary condition.
The eigenvalue equation is
\begin{eqnarray}
q_{j+1}-q_{j} + \frac{\beta}{2N}(q_{j+1}+q_{j}) & = & \epsilon p_{j+1},\\
p_{j}-p_{j+1} + \frac{\beta}{2N}(p_{j}+p_{j+1}) & = & \epsilon q_{j}
\;\;\; (1\le j\le N-1).
\end{eqnarray}
\begin{eqnarray}
  q_{1} + q_{N} + \frac{\beta}{2N}(q_{1}-q_{N}) & = & \epsilon p_{1},\\
  p_{N} + p_{1} + \frac{\beta}{2N}(p_{N}-p_{1}) & = & \epsilon q_{N} 
\end{eqnarray}
We assume the solutions of eigenvalue equations as
\begin{equation}
\left(
\begin{array}{c}
p_{j}\\
q_{j}
\end{array}
\right) = 
\left(
\begin{array}{c}
p \\
q
\end{array}
\right) 
\exp \left(i\frac{(2r+1)\pi}{N}j\right)
\;\;\;(r=0,1,...,N-1).
\end{equation}
Then the eigenvalue equation becomes
\begin{equation}
\left(
\begin{array}{cc}
0 & 1-e^{-i\frac{\pi (2r+1)}{N}}+
\frac{\beta}{2N}(1+e^{-i\frac{\pi (2r+1)}{N}}) \\
1-e^{i\frac{\pi (2r+1)}{N}}+\frac{\beta}{2N}(1+e^{i\frac{\pi (2r+1)}{N}}) & 0
\end{array}
\right) 
\left(
\begin{array}{c}
p\\
q
\end{array}
\right) =
\epsilon
\left( 
\begin{array}{c}
p\\
q
\end{array}
\right),
\end{equation}
\begin{equation}
\epsilon^{2} - 
\left(\frac{\beta^{2}}{N^{2}}\cos^{2}\frac{\pi (2r+1)}{2N} +
4\sin^{2}\frac{\pi (2r+1)}{2N}\right)
= 0.
\label{heq}
\end{equation}
Therefore
\begin{eqnarray}
|\det \Dslash_{N}| & = &  \prod_{r=0}^{N-1}\left\{
\left(\frac{\beta}{N}\right)^{2}\cos^{2}\frac{\pi (2r+1)}{2N}
+ 4\sin^{2}\frac{\pi (2r+1)}{2N}\right\},\\
& = & 
\prod_{r=0}^{N-1}\left(4\sin^{2}\frac{\pi (2r+1)}{2N}\right)
\prod_{r=0}^{N-1}\left(1 + \frac{\beta^{2}}{4N^{2}}
\cot^{2}\frac{\pi (2r+1)}{2N}\right).
\end{eqnarray}
\begin{eqnarray}
\lim_{N\rightarrow\infty}|\det \Dslash_{N}| & = &
4\prod_{r=0}^{\infty}\left\{1 + \left(\frac{\beta}{\pi (2r+1)}\right)^{2}
\right\}^{2},\\
& = & 4\cosh^{2}\frac{\beta}{2}.
\end{eqnarray}
Here we used (\ref{d_norm2}) and (\ref{cosh}).
Since the inverse hyperbolic type can be deformed continuously to
elliptic type ($\beta\rightarrow$0), the Maslov index of this part
is the same as the elliptic type. Therefore Maslov index of this
part is 1.

\subsection{Loxodromic type}

The normal Hamiltonian of this type is
\begin{equation}
H = \alpha (p_{1}q_{2}-p_{2}q_{1}) -
    \beta  (p_{1}q_{1}+p_{2}q_{2}).
\end{equation}
The eigenvalue equation is
\begin{eqnarray}
q_{1,j}-q_{1,j-1} + \frac{\beta}{2N}(q_{1,j}+q_{1,j-1}) 
- \frac{\alpha}{2N}(q_{2,j}+q_{2,j-1}) & = & \epsilon p_{1,j},\\
q_{2,j}-q_{2,j-1} + \frac{\alpha}{2N}(q_{1,j}+q_{1,j-1})
+ \frac{\beta}{2N}(q_{2,j}+q_{2,j-1}) & = & \epsilon p_{2,j},\\
p_{1,j}-p_{2,j+1} + \frac{\beta}{2N}(p_{1,j}+p_{1,j+1}) 
+ \frac{\alpha}{2N}(p_{2,j}+p_{2,j+1}) & = & \epsilon q_{1,j},\\
p_{2,j}-p_{2,j+1} - \frac{\alpha}{2N}(p_{1,j}+p_{2,j+1})
+ \frac{\beta}{2N}(p_{2,j}+p_{2,j+1}) & = & \epsilon p_{2,j}.
\end{eqnarray}
If we assume the forms of eigenvectors of $\Dslash_{N}$ as
\begin{equation}
\left(
\begin{array}{c}
p_{1,j}\\p_{2,j}\\
q_{1,j}\\q_{2,j}
\end{array}
\right) = 
\left(
\begin{array}{c}
p_{1}\\p_{2}\\
q_{1}\\q_{2}
\end{array}
\right) 
\exp \left(i\frac{2\pi r}{N}j\right)
\;\;\;(r=0,1,...,N-1),
\end{equation}
the eigenvalue equation become
\begin{equation}
\left(\begin{array}{cc}
0 & A_{r}\\
A_{r}^{\dagger} & 0
\end{array}\right)
\left(
\begin{array}{c}
p_{1}\\ p_{2}\\
q_{1}\\ q_{2}
\end{array}
\right) =
\epsilon
\left( 
\begin{array}{c}
p_{1}\\ p_{2}\\
q_{1}\\ q_{2}
\end{array}
\right),
\end{equation}
\begin{equation}
A_{r} = \left(\begin{array}{cc}
1-\eminus + \frac{\beta}{2N}(1+\eminus) & 
-\frac{\alpha}{2N} (1+\eminus) \\
\frac{\alpha}{2N} (1+\eminus) & 
1-\eminus + \frac{\beta}{2N}(1+\eminus)
\end{array}\right),
\end{equation}
\begin{equation}
\det (\epsilon^{2}I -  A_{r}A_{r}^{\dagger}) = 0.
\label{leq}
\end{equation}
The explicit form of $A_{r}A_{r}^{\dagger}$ is 
\begin{equation}
A_{r}A_{r}^{\dagger} = \left(
\begin{array}{cc}
\frac{\alpha^{2}+\beta^{2}}{N^{2}}\cos^{2}\frac{\pi r}{N} 
+ 4\sin^{2}\frac{\pi r}{N}&
4i \frac{\alpha}{N}\sin\frac{\pi r}{N}\cos\frac{\pi r}{N}\\
-4i \frac{\alpha}{N}\sin\frac{\pi r}{N}\cos\frac{\pi r}{N} &
\frac{\alpha^{2}+\beta^{2}}{N^{2}}\cos\frac{\pi r}{N}
+ 4\sin^{2}\frac{\pi r}{N}
\end{array}
\right).
\end{equation}
Therefore the eigenvalue equation becomes
\begin{equation}
\epsilon^{4} - 
2\left(\frac{\alpha^{2}+\beta^{2}}{N^{2}}\cos^{2}\frac{\pi r}{N}
+4\sin^{2}\frac{\pi r}{N}\right)\epsilon^{2} +
\det A_{r}A_{r}^{\dagger} = 0,
\end{equation}
\begin{equation}
\det A_{r}A_{r}^{\dagger} =
\left(4\sin^{2}\frac{\pi r}{N} + 
\frac{\beta^{2}-\alpha^{2}}{N^{2}}\cos^{2}\frac{\pi r}{N}\right)^{2}
+ \frac{4\alpha^{2}\beta^{2}}{N^{4}}\cos^{4}\frac{\pi r}{N}
\end{equation}
Hence
\begin{eqnarray}
|\det \Dslash_{N}| & = & \prod_{r=0}^{N-1}
\det A_{r}A_{r}^{\dagger},\\
& = & \left(\frac{\alpha^{2}+\beta^{2}}{N^{2}}\right)^{2}\prod_{r=1}^{N-1}
\left(2\sin\frac{\pi r}{N}\right)^{4}
\left\{\left(1+\frac{\beta^{2}-\alpha^{2}}{4N^{4}}\cot^{2}\frac{\pi r}{N}
\right)^{2} 
+ \frac{\alpha^{2}\beta^{2}}{4N^{4}}\cot^{4}\frac{\pi r}{N}\right\},\\
& = &
(\alpha^{2}+\beta^{2})^{2}\prod_{r=1}^{N-1}
\left\{\left(1+\frac{\beta^{2}-\alpha^{2}}{4N^{4}}\cot^{2}\frac{\pi r}{N}
\right)^{2} 
+ \frac{\alpha^{2}\beta^{2}}{4N^{4}}\cot^{4}\frac{\pi r}{N}\right\}.
\end{eqnarray}
\begin{eqnarray}
\lim_{N\rightarrow\infty}|\det \Dslash_{N}| & = &
(\alpha^{2}+\beta^{2})^{2}
\prod_{r=1}^{\infty}\left\{
\left(1 + \frac{\beta^{2}-\alpha^{2}}{(2\pi r)^{2}}\right)^{2} + 
\frac{(2\alpha\beta)^{2}}{(2\pi r)^{4}}\right\}^{2},\\
& = & (2\pi)^{4}(\gamma^{4}+\delta^{4})
\prod_{r=1}^{\infty}
\left(1 - \frac{\gamma^{2}}{r^{2}}\right)^{4}
\prod_{r=1}^{\infty}\left\{
1 + \frac{\delta^{4}}
{(r^{2} - \gamma^{2})^{2}}\right\}^{2}.
\end{eqnarray}
where
\begin{eqnarray}
\gamma^{2} & = & \frac{\alpha^{2} - \beta^{2}}{(2\pi)^{2}},\\
\delta^{2} & = & \frac{2\alpha\beta}{(2\pi)^{2}}. 
\end{eqnarray}
Thus we obtain the result
\begin{eqnarray}
|\det \Dslash_{N}| 
& = & 4 (\cosh \beta - \cos \alpha)^{2},
\end{eqnarray}
using the formulas (\ref{sin}) and (\ref{sp}).

Since the eigenvalue equation (\ref{leq}) always has 
two positive and two negative eigenvalues, the Maslov
index of this part is 0.

\subsection{Parabolic type}

The Hamiltonian is 
\begin{equation}
H = \frac{\gamma}{2}q^{2},
\end{equation}
and the eigenvalue equation is
\begin{equation}
\left(
\begin{array}{cc}
0 & 1 - e^{-i\frac{2\pi r}{N}} \\
1 - e^{i\frac{2\pi r}{N}} & 
-\frac{\gamma}{4N}(2 + e^{i\frac{2\pi r}{N}} 
+ e^{-i\frac{2\pi r}{N}})
\end{array}
\right) 
\left(
\begin{array}{c}
p\\
q
\end{array}
\right) =
\epsilon
\left( 
\begin{array}{c}
p\\
q
\end{array}
\right),
\end{equation}
\begin{equation}
\epsilon^{2} + 
\frac{\gamma}{N} \cos^{2}\frac{\pi r}{N}\epsilon 
- 4\sin^{2}\frac{\pi r}{N}
= 0.
\label{peq}
\end{equation}
If $r\ne 0$, the eigenvalues of (\ref{peq}) are not equal
to zero, and the determinant of this part is
\begin{eqnarray}
\prod_{r=1}^{N}
\left(-4\sin^{2}\frac{\pi r}{N}\right) & = & 
(-1)^{N-1}N^{2}.
\end{eqnarray} 
Since the two solutions of (\ref{peq}) have opposite signs, these
eigenvalues don't contribute to the Maslov index.

If $r=0$, the eigenvalues are
\begin{equation}
\epsilon = 0, - \frac{\gamma}{N}
\end{equation}
and the Maslov index of this part is $\frac{1}{2}{\rm sgn}\gamma$.
The normalized eigenvector corresponding to these eigenvalues are
\begin{equation}
\left(\begin{array}{c}p_{j}\\q_{j}\end{array}\right)
= \frac{1}{\sqrt{N}}
\left(\begin{array}{c}1 \\ 0\end{array}\right),
\frac{1}{\sqrt{N}}
\left(\begin{array}{c}0 \\ 1\end{array}\right).
\end{equation}
The volume factor of the zero-mode measured by these vectors
are $V_{N} = \sqrt{N}V$, where $V$ is the volume factor by the normal
measure in the phase space. 
Therefore the contribution from this part is
\begin{equation}
\int \frac{d\bX_{N}}{(2\pi\hbar)^{N}} 
\exp \left[\frac{i}{2\hbar}\bX_{N}^{T}\Dslash\bX_{N}\right] =
\frac{V_{N}}{\sqrt{2\pi\hbar iN\gamma}} =
\frac{V}{\sqrt{2\pi\hbar i\gamma}}.
\end{equation}

%% file: formula.tex
\section{Useful formulas}
\begin{equation}
\prod_{n=1}^{\infty}\left(1-\frac{x^{2}}{n^{2}}\right)
= \frac{\sin\pi x}{\pi x}
\label{sin}
\end{equation}

\begin{equation}
\prod_{n=1}^{\infty}\left(1+\frac{x^{2}}{n^{2}}\right)
= \frac{\sinh\pi x}{\pi x}
\label{sinh}
\end{equation}

\begin{equation}
\prod_{n=0}^{\infty}\left(1+\frac{x^{2}}{(2n+1)^{2}}\right)
= \cosh \frac{\pi x}{2}
\label{cosh}
\end{equation}

\begin{equation}
\prod_{n=-\infty}^{\infty}\left(1+\frac{x^{2}}{(a+2n\pi)^{2}}\right)
= \frac{\cosh x - \cos a}{1 - \cos a}
\label{chcos}
\end{equation}

\begin{equation}
\prod_{n=1}^{\infty}\left(1 + \frac{x^{4}}{(n^{2}-a^{2})^{2}}\right)
= \frac{a^{2}}{\sqrt{x^{4}+a^{4}}}
\frac{\cosh 2I - \cos 2R}{2\sin^{2}\pi a}
\label{sp}
\end{equation}
\begin{eqnarray}
R & = & {\rm Re}\; \pi\sqrt{a^{2} + ix^{2}}\\
I & = & {\rm Im}\; \pi\sqrt{a^{2} + ix^{2}}
\end{eqnarray}

\begin{equation}
\prod_{r=1}^{N-1}\sin \frac{\pi r}{N} = \frac{N}{2^{N-1}}
\label{d_norm}
\end{equation}

\begin{equation}
\prod_{r=0}^{N-1}\sin\frac{(2r+1)\pi}{2N} = \frac{1}{2^{N-1}}
\label{d_norm2}
\end{equation}

%% file: main.bbl
\begin{thebibliography}{99}
\bibitem{gutzwiller}
        M.C.Gutzwiller,{\em Chaos in Classical and Quantum Mechanics},
        (Springer Verlag,1990). 
\bibitem{robbins}
        J.M.Robbins,Nonlinearity {\bf 4}(1991),343.
\bibitem{brack}
        M.Brack and S.R.Jain,Phys.Rev.{\bf A 51}(1995),3462.
\bibitem{kashiwa} 
	T.Kashiwa, Y. Ohnuki and M. Suzuki,
        {\it Path Integral Methods}, Clarendon Press, Oxford (1997).
\bibitem{levit2}
        S.Levit,K.M\"{o}hring,U.Smilansky and T.Dreyfus,
        Ann.Physics {\bf 114}(1978),223.
\bibitem{sakita}
        B.Sakita and K.Kikkawa, {\em Quantum Many Body Problems
in terms of path integral methods},(Iwanami Shoten,1986,in Japanese).
\bibitem{kuratsuji}
        H.Kuratsuji,Phys.Lett.{\bf 108A}(1985)139.
\bibitem{creagh}
        S.C.Creagh and R.G.Littlejohn,Phys.Rev.{\bf A 44}(1991),836.
\bibitem{sugita} 
	A. Sugita, Phys. Lett. {\bf A 266}(2000) 321.
\bibitem{witten} 
	E. Witten, Phys. Lett. {\bf B 117} (1982), 324.
\bibitem{elitzure}
	S. Elitzure, E. Rabinovici, Y. Frishman and A. Schwimmer, 
	Nucl. Phys. {\bf B273} (1986), 93.
\bibitem{littlejohn}
	R. G. Littlejohn, Phys. Rep. {\bf 138} (1986), 193.
\bibitem{arnold}
	V.I.Arnold, {\it Mathematical methods of classical Mechanics},
	(Springer Verlag, 1990).
\end{thebibliography}
